\documentclass[12pt,preprint]{aastex}

\usepackage{natbib}
\usepackage{lscape}

\lefthead{Christopher et al.}
\righthead{\HCN~and \HCO~CND Observations}
\def\sgra {$\rm {Sgr~A^{\rm {*}}}$}
\def\kms {km $\rm {s^{-1}}$}
\def\cmcube {$\rm {cm^{-3}}$}
\def\HCNF {HCN (J=1-0)}
\def\HCN {HCN}
\def\HCOF {HC$\rm O^{\rm +}$~(J=1-0)}
\def\HCO {HC$\rm O^{\rm +}$}
\def\jybeam {Jy $\rm {beam}^{-1}$}
\def\ta{$\tau_{\rm {HCN}} =4$}
\def\mdot {$\rm{M}_{\odot}$}

\begin{document}

\title{\HCN~and
  \HCO~Observations of the Galactic Circumnuclear Disk}
\author{M.H. Christopher}
\affil{Astronomy Department, California Institute of Technology,
  Pasadena, CA 91125; mc@astro.caltech.edu}

\author{N.Z. Scoville}
\affil{Astronomy Department, California Institute of Technology,
  Pasadena, CA 91125; nzs@astro.caltech.edu}

\author{S.R. Stolovy}
\affil{Spitzer Science Center, California Institute of Technology,
  Pasadena, CA 91125; stolovy@ipac.caltech.edu}

\and

\author{Min S. Yun}
\affil{Astronomy Department, University of Massachusetts, Amherst, MA
  01003; myun@astro.umass.edu}
\slugcomment{Received 2004 February 11; Accepted 2004 December 3}

\begin{abstract}
We present high spatial resolution (5.1\arcsec~$\times$~2.7\arcsec)
Owens Valley Radio Observatory (OVRO) millimeter array observations of
\HCNF~and \HCOF~emission in the inner 3 pc of the Galaxy (0.04~pc~$\sim$~ 
1\arcsec).  The \HCN~and
\HCO~emission of the Circumnuclear Disk (CND) is distributed in a 
well-defined ring with a
peak at a radius of 1.6 pc.  The observed
radial velocities are generally consistent with rotation at $\sim110$~\kms~(except along the western edge 
of the CND). The \HCO/\HCN~emission ratio is typically~$\sim$~0.4 but
with significant variations.  The variations in the 
\HCO/\HCN~emission and absorption
ratios can be attributed to greater abundances of \HCO~in lower
density regions both within the CND and along the line-of-sight. The 
\HCN~emission is well correlated with the H$_2$ emission at 2.12 $\mu$m
both in 
the main emission lobes of the CND and also in four H$_2$ and \HCN~filaments. Multiple areas of interaction
between the ionized gas and the CND are also seen -- the western
arm of the minispiral is spatially and kinematically consistent with
being the ionized inner edge of the CND, and the northern arm
may be connected to the CND northeast extension.  

With the enhanced spatial resolution of the \HCN~maps, we resolve
numerous dense molecular gas cores within the CND with characteristic
diameter~$\sim~7\arcsec$~(0.25 pc).  For 26 of the more isolated
cores, we have measured sizes, velocity widths, and integrated
fluxes.  From these properties we estimated 3 masses for each core: a
virial mass assuming the cores are gravitationally bound, an optically
thick mass from observed column densities of \HCN, and a lower limit mass
assuming the \HCN~emission is optically thin and shock-excitation is negligible.  The virial and optically thick
masses are in good agreement with a typical mass of
(2-3)$\times10^4$~\mdot~and a total CND gas mass of 10$^6$~\mdot.  The
internal densities implied by these core masses (assuming a uniform
density distribution for each core) are on average
(3-4)$\times10^7$\cmcube.  The core densities are high enough
to be stable against tidal disruption from \sgra~and the central
stellar concentration.  This tidal stability suggests a longer
lifetime for the CND. The high densities and masses within the cores
might support star formation either in the CND itself or within a core
infalling towards the inner parsec, thus providing a mechanism for the
formation of the young stellar population observed in the inner
arcseconds of the galaxy.
\end{abstract}


\section{Introduction}
At a distance of only 8 kpc~\citep{Reid1993}, the Galactic Center
presents a unique opportunity to study in exquisite detail a
supermassive black hole and its interactions with the surrounding
medium. Extensive
observations of the inner 3 parsecs of the Galaxy have revealed an
extremely complicated, highly interacting region. Found within these 
inner
parsecs are: a $4.0\times10^6~\rm {M_{\odot}}$ black hole
~\citep{Ghez2003a} coincident
with the non-thermal radio source \sgra~\citep{Backer1999, Reid1999};
a dense stellar population, including a central cluster of bright, 
young (less than a few Myr)
  emission-line stars~\citep{Morris1996, Krabbe1995}; X-ray emission,
  both in hundreds of point sources and in diffuse
  emission~\citep{Muno2003}; streamers of ionized gas and dust, 
dubbed the ``minispiral''~\citep{Ekers1983, Lo1983,
  Lacy1991, Scoville2003}; and a clumped ring of molecular gas, the
  Circumnuclear Disk (CND), surrounding \sgra, the minispiral, and 
the central star
  cluster~\citep[e.g.][]{Becklin1982,Gusten1987, Wright2001}.  

The CND was first detected by \citet{Becklin1982} as double-lobed 
emission at $50\mu\rm {m}$ and $100\mu\rm {m}$; it has subsequently 
been observed extensively at radio to infrared wavelengths
(see~\citet{Vollmer2001b}).
These observations indicate that the CND is a ring-like structure of
molecular gas and dust with an inclination of $50\arcdeg~\rm{-}~75\arcdeg$
and a position angle of~$\sim 25\arcdeg$~\citep{Jackson1993}.  The
ring is nearly complete in \HCN~but has noticeable gaps, especially to the north~\citep{Wright2001}.  The inner radius of the 
CND is
well-defined at $\sim 1.5$~pc (deprojected).  Early single dish 
observations traced \HCN, CO, and \HCO~out to more than 7
pc~\citep{Genzel1985, Lugten1986, Serabyn1986, Gusten1987}, but a
recent interferometric study suggests an outer edge at 3 - 4
pc~\citep{Wright2001}. The CND has a thickness of $\sim 0.4$~pc at 
the inner edge~\citep{Gusten1987,
  Jackson1993} and may expand to 2 pc at larger distances from
\sgra~\citep{Vollmer2001b}.  CND kinematics are consistent with rotation at
110~\kms~\citep{Marr1993} with deviations in some regions, 
most notably to the west.  These deviations are consistent with a
warped disk created from the collision of
two large clouds~\citep{Gusten1987} or a ring of multiple dynamically 
distinct streamers~\citep{Jackson1993}.  

To understand the behavior of the CND in the complexity of the
Galactic Center, and in particular to evaluate its stability, lifetime,
and potential for star formation, it is critical to constrain
the physical conditions of the CND molecular gas.
Previous
observations have found a total mass within the CND of only
$\sim$~$10^4$~M$_\odot$~\citep{Genzel1985} and indicated 
densities of $\sim$~$10^5$~-~$10^6$~\cmcube~\citep{Marr1993}. These
densities are sufficiently low that the gas cores would be unstable to tidal 
disruption from \sgra~and the central stellar population.
However,~\citet{Jackson1993} calculate densities within the CND cores of
10$^6$~-~$10^8$~\cmcube~from models of the \HCN~emission, noting that
such densities would significantly increase the expected core lifetime.  They do point out that modeling of other molecular and
atomic tracers yields lower densities in keeping with the values
of~\citet{Marr1993}.   Here we
present new observations of the CND in~\HCNF~and~\HCOF~at a 
resolution of
5.1\arcsec~$\times$~2.7\arcsec, a factor of 2-3 gain in spatial
resolution over previous studies.  With this enhanced resolution, we 
resolve
substantial core structure within the CND and calculate typical core
densities of $10^7$~-~$10^8$~\cmcube, in agreement with the results
of~\citet{Jackson1993}.  In addition,, we measure typical gas core
masses of a few$\times~10^4$~M$_\odot$, larger than previous estimates.  

\section{Observations}\label{sec:observations}
The observations described here were taken from 1999 November to
2000 April at the Owens Valley Radio Observatory (OVRO) millimeter
array.  We obtained 10 tracks in
  four configurations (Table~\ref{tab:tracks}), resulting in a
  naturally-weighted synthesized beam FWHM of
$5.1\arcsec \times 2.7\arcsec$. The
  observations used a 10 pointing mosaic to cover a 
$120\arcsec~\times210\arcsec$~($4.8 \times 8.4~\rm{pc}$) region
  centered on \sgra.  Because of the low
declination (-29\arcdeg) of \sgra, each track was only 3-4 hr in
length.  In total, 115 minutes
were spent on each pointing during the observations,
resulting in a $1\sigma$~noise rms of $\sim 30~\rm {mJy}~\rm 
{beam}^{-1}$
in each 6.7~\kms~spectral channel and $6~\rm {mJy}~\rm {beam}^{-1}$
in the 1 GHz wide continuum band.

We simultaneously imaged the \HCNF~and \HCOF~transitions in the upper 
sideband at 88.63 and
89.19 GHz, respectively.  \HCNF~and \HCOF~are both
tracers of high density ($>10^5$-$10^6$~\cmcube) molecular gas and 
are therefore ideal molecules for
probing the dense CND cores.  
 Each line was observed in 64 channels of 2 MHz (6.8~\kms) width centered on
V$_{\rm {LSR}}~=~0$~\kms.  The total velocity range was -213 to
206~\kms, easily covering the -150 to 150~\kms~region where emission
was detected in previous observations.

In addition, we obtained a single track in the L configuration of
the H$^{13}$CN line at 86.34 GHz.  We observed the two pointings
covering the northeastern and southwestern emission lobes of the CND and
used these observations to estimate the \HCN~optical depth
($\tau_{\rm{HCN}}$) for the CND (\S\ref{sec:cmass}).
 
A bright quasar was observed twice in each
track for passband and flux calibration. The quasar fluxes, derived
from comparisons with observations of Uranus and Neptune in nearby tracks, varied slowly over the six 
months of
the observation by as much as 20\% (Table~\ref{tab:tracks}), but this
variation was accounted for in the individual flux
calibration of each track.  We estimate a total uncertainty in the
flux measurements of $\sim$~10\%.  NRAO 530 was observed every 15 
minutes
for gain calibration, and based on the NRAO 530 observations, periods 
of low
coherence on each baseline were removed. All of the calibration and
editing was done with MMA~\citep{Scoville1993}, while all cleaning 
and mapping used the Miriad software package~\citep{Sault1995}.  

The dominant noise in these
interferometer images is misplaced flux of bright emission peaks
resulting from residual atmospheric and instrumental phase
fluctuations.  Using the strong signal from the continuum source and
the closure phase relationship (see~\citet{Cornwell1981}), the random
phase fluctuations are corrected.  Such iterative phase self-calibration was performed on the five pointings
containing \sgra~in the primary beam
(diameter~$\sim 90\arcsec$ for 89 GHz) using line free
channels to produce the continuum model by cleaning down to the first
negative clean component.   The continuum model is
dominated by point source emission from \sgra.  The
typical time interval for self-calibration was 30 minutes,
corresponding to the period between repetitions of the same map position.
The self-calibration reduced the noise in the images by 20 - 30$\%$
and 
increased the total recovered flux for \sgra by $\simeq$~50\%. Since
the self-calibration was based on line-free channels and contained
only \sgra in the model, 
 the CND emission line fluxes were increased negligibly by the self-calibration (typically by only 5\%
with a maximum increase of 15\%).  As noted above, the self-calibration was used 
to reduced the noise due to mis-calibrated and/or scattered flux of \sgra. Therefore, the lack of self-calibration on the five positions at the northern and southern edges of
the field without \sgra~in the primary beam did not significantly
affect our determination of the CND line flux.

\begin{figure}[htb]
\begin{center}
\epsscale{0.7}
\plotone{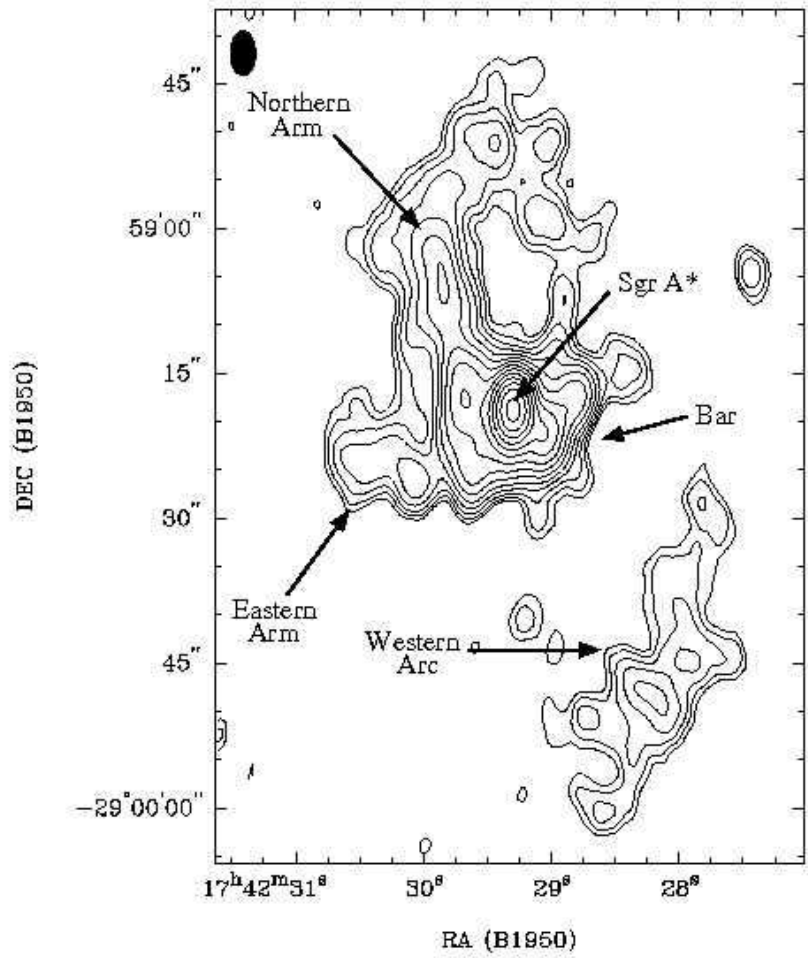}
\caption{Continuum emission at 85.9 GHz from \sgra~and the minispiral. The
  contours start at 15 m\jybeam~and
  increase by factors of $2^{\frac{1}{2}}$.  The main components of the minispiral
  are detected in the continuum image and labeled above.  The
  beam FWHM is indicated by the filled oval. }
\label{fig:continuum}
\end{center}
\end{figure}

The 10 pointings were mosaicked in the UV plane using the Miriad
invert command with the ``systemp'' option, which gives low
statistical weights to visibilities with high
system temperatures.  The maps were cleaned to a 1.5$\sigma$~level in 
each
individual channel using Miriad routine mossdi2, which is designed
for mosaicked observations and cleans to a given level of the 
theoretical noise at each point in a mosaic
rather than to a set flux level. We then subtracted the 85.91 GHz
continuum map (Fig.~\ref{fig:continuum}) in the image plane from each
channel map to produce the line emission maps.

\section{Results}\label{sec:results}

\subsection{\sgra~Continuum Observations}\label{sec:continuum}

The 85.9 GHz continuum map (Fig.~\ref{fig:continuum}) traces the
ionized thermal emission of the minispiral and the non-thermal
emission of \sgra.  We adopt a location for \sgra~at the peak of the
85.9 GHz continuum emission (17:42:29.3, -28:59:18.7, B1950).  All the features of the
minispiral are detected at $>3\sigma$ level and are labeled in the 
figure.  The peak
continuum emission is 1.6~\jybeam~in the
$5.1\arcsec~\times~2.7\arcsec$~beam. \citet{Wright2001} measure a
peak flux of 2.14~\jybeam~for a $13\arcsec~\times~4\arcsec$~beam.  
Convolving our observations to
match their resolution, we obtain a peak flux of
2.02~\jybeam, consistent with their results.  For an aperture of 30\arcsec~radius centered on \sgra, our 
map contains 5.38 Jy of continuum emission, compared with 8.9 Jy from~\citet{Wright2001}. This 
difference is likely due 
to the shorter baselines available in the Berkeley-Illinois-Maryland
Association (BIMA) data, which enable 
detection of a greater fraction of the spatially extended components
of the thermal emission.  Tracing recovered flux as a function of
distance from \sgra, we find very little additional flux recovered in
apertures larger than 30-40\arcsec.  Nobeyama
45-m single dish observations~\citep{Tsuboi1988}, free of the missing
short spacings problem, likewise observe little emission beyond a
30-40\arcsec~radius from \sgra~and measure 27 Jy within 60$\arcsec$
of \sgra~\citep{Wright1993}, compared with $\sim 5.4$~Jy from our
observations, suggesting we recover $\sim$~20\% of the total 3mm 
continuum flux.  

\subsection{Line-of-sight Absorption}\label{sec:sgraspectrum}

Absorption by molecular gas along the line of sight to the Galactic
Center can significantly effect the observed \HCN~and \HCO~emission at some velocities.  Figure~\ref{fig:sgraspectrum} shows 
\HCN~and \HCO~spectra extracted from a 4\arcsec~diameter region 
centered
on \sgra.  Both spectra show broad
absorption from -80~to
+70~\kms~from Galactic center gas and narrower absorption features 
at -135, -50,
and -30~\kms~due to foreground Galactic arms 
at radii of 250 pc~\citep{Scoville1972, Liszt1978, Bieging1980}, 3
kpc~\citep{Oort1977}, and 4 kpc~\citep{Menon1970} respectively.

Although the velocities of the \HCN~and \HCO~absorption features are generally 
similar, the \HCO~absorption is deeper than that of the \HCN~at
-135~and 20~-~40~\kms.  These deeper \HCO~features suggest
regions of lower density gas along the line of sight where the fractional
 abundance of \HCO~has been enhanced because of deeper penetration by cosmic
rays~\citep{Marr1993}. 

\subsection{\HCN~and~\HCO~Emission from the CND}\label{sec:moment0} 
The integrated
\HCN~and~\HCO~intensity maps (Figs.~\ref{fig:hcnmoment0}
and~\ref{fig:hcomoment0}) clearly show a ring-like CND with a gap 
in
emission to the north of \sgra~and reduced emission to the east.  The
overall CND gas distribution is 
similar
to that seen in previous imaging, but these observations detect more
significant emission to the east of \sgra~and because of the enhanced spatial resolution reveal much greater internal
structure within the CND ring.  The \HCN~flux is on average 2 - 3 
times the \HCO~flux (see
\S\ref{sec:hcnvshco}).  The CND flux distribution as a function of
distance from \sgra~(deprojected assuming an 
inclination angle of 60\arcdeg) has a
sharp inner edge at 40$\arcsec$~(1.6 pc), an equally steep
decline around 50$\arcsec$~(2.0 pc), followed by a slowly declining, non-zero flux out past 150$\arcsec$ (Fig.~\ref{fig:fluxversusdistance}). The sharp inner and
outer edges strongly suggest the CND has a ring-like
morphology~\citep{Wright2001}.  Despite its ring-like nature, we
nevertheless maintain the traditional nomenclature of circumnuclear
disk (CND).

Channel maps of the \HCN~and \HCO~emission binned in 27~\kms~(four 
channel)
intervals are shown in Figures~\ref{fig:hcnchannels} 
and~\ref{fig:hcochannels}. The rotation of the CND is clearly evident, as the 
dominant
negative velocity emission arises southwest of \sgra~and loops around
to the northeast of \sgra, where the
emission is primarily at positive velocities. The channel maps
also clearly show locations with multiple velocity components.
Perhaps the most prominent example is 20$\arcsec$~west and
20$\arcsec$~south of \sgra, where significant emission is seen around $-40$~and
70~-~100~\kms.

The rotation of the CND is additionally apparent in the
flux-weighted average velocity map
(Fig.~\ref{fig:momentone}).  The velocity structure is continuous
around the ring with the greatest negative (approaching)
velocity near the southwestern emission lobe and the greatest
positive (receding) velocity near the northeastern emission lobe.
The sharp
velocity gradient along the western side of the CND is a result of the
spatial coincidence of -50~and 70 - 110~\kms~emission and the
presence of strong absorption at velocities near 0~\kms.
The region to the east of \sgra~has a velocity near
0~\kms~and also suffers from large amounts of absorption, which explains the reduction in observed emission.  We note that the
primary velocities expected in the northern gap ($\sim~50 -70~$\kms)
do not suffer from severe line of sight absorption, thereby suggesting
a true dearth of molecular emission in the northern gap.

\subsection{Comparing \HCO~to \HCN~Emission in the 
CND}\label{sec:hcnvshco}

The \HCO/\HCN~abundance ratio is~$\sim$~1 for typical Galactic
molecular clouds~\citep{Nyman1983,Vogel1983,Blake1987}.  Previous
observations of the CND found that the \HCO~emission traces the same 
regions as the \HCN~emission,
 but with \HCO/\HCN~as
 low as 0.06 - 0.2~\citep{Marr1993}. As seen in the
 integrated maps (Figs.~\ref{fig:hcnmoment0} 
and~\ref{fig:hcomoment0}), the relative
 strength of \HCN~versus \HCO~emission in our observations
 is largely consistent with these previous measurements.  For 
locations with
 $>3\sigma$ emission in both \HCN~and \HCO, the mean \HCO/\HCN~ratio 
is
 0.74. However, this result is misleading because we are only
 considering points with 3$\sigma$ emission in both tracers.  There 
are almost
 3 times as many locations with $>3\sigma$ \HCN~emission without
 3$\sigma$~\HCO~emission as there are locations with 3$\sigma$
 \HCO~emission and $<3\sigma$~\HCN~emission.  Therefore, the
 true {\it average} \HCO/\HCN~emission ratio is significantly lower.  
Comparing the ratio of total \HCO~to total \HCN~emission in
 each channel, we find a typical channel \HCO/\HCN~ratio of 0.4 (Fig.~\ref{fig:hcnhcoratio}).

\begin{figure}[htb]
\begin{center}
\epsscale{1.0}
\plotone{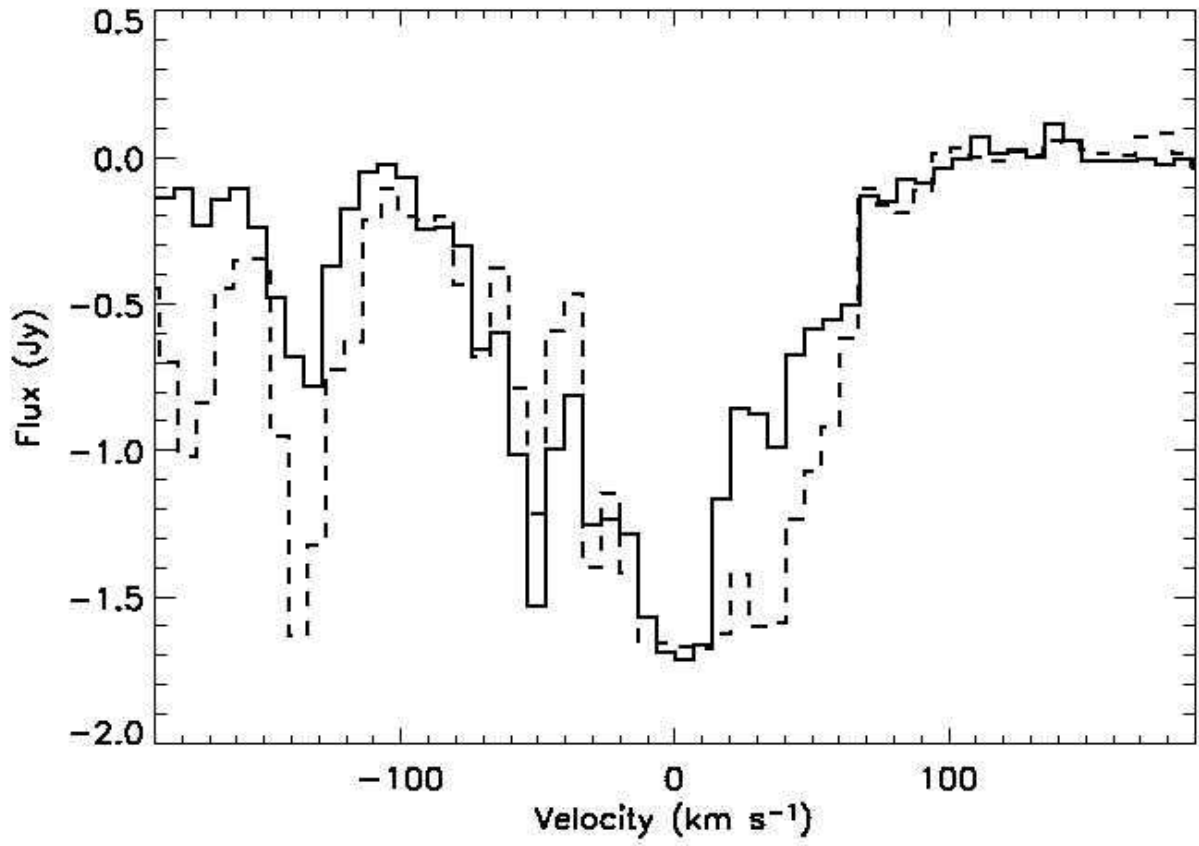}
\caption{\HCN~(solid) and \HCO~(dashed) spectra extracted from a
  4\arcsec~diameter region centered on \sgra.  Absorption features at
  -135, -50, and -30 \kms, as well as an absorption trough centered around
  0 \kms, are detected with varying depths in \HCN~and \HCO, perhaps
  indicating differing densities of absorbing material.}
\label{fig:sgraspectrum}
\end{center}
\end{figure}

Examining the variation of the \HCO/\HCN~ratio with velocity we find that
the lowest ratios are found in the channels corresponding to the
 northeastern and southwestern emission lobes of the CND with the highest
 ratios around 0~\kms~(Fig.~\ref{fig:hcnhcoratio}).   This region of
 enhanced \HCO~emission relative to the \HCN~emission corresponds roughly to the
 broad absorption trough around
 0~\kms~(Fig.~\ref{fig:sgraspectrum}).  While there are velocity
 ranges in which~\HCO~absorption on \sgra~is more significant than
 \HCN~absorption, especially around -140~and 20-40\kms
 (Fig.~\ref{fig:sgraspectrum}), the \HCO/\HCN~ratio in these regions
 is neither
 significantly enhanced nor reduced from the values in
 surrounding channels, suggesting that the differential line of sight
 absorption between \HCN~and \HCO~does not significantly affect the analysis of the
 \HCO/\HCN~ratio in each channel.  As we will see in
 \S\ref{sec:coreproperties}, the southwestern and northeastern lobes
 include some of the most massive cores, while cores in our sample
 that are found at velocities around 0~\kms~tend to be the least
 massive.  This suggests a trend where the lowest \HCO/\HCN~ratios
 correspond to high mass regions while the highest \HCO/\HCN~ratios are
 found in the lower mass regions.  We estimate the mass of each core
 (see \S\ref{sec:masses}) in three independent ways.  Two of these
 techniques do have a dependence upon the level of the \HCN~emission
 (or the abundance of \HCN~relative to H$_2$), which potentially biases us towards high mass cores
 in regions with high \HCN~flux (and thus low \HCO/\HCN~ratios).
 However, the third tracer, requiring an assumption that the cores
 are virialized, is independent of the
 observed \HCN~flux (and as we shall see provides density and mass
 estimates in good agreement with estimates from assuming the cores to
 be optically thick to \HCN~emission), thus strengthening the support for the
 correlation of low
 \HCO/\HCN~ratios in high mass regions.  This correlation could result from the 
CND having enhanced HCN emission in more massive regions because of increased shock processing of the 
gas (perhaps in core-core collisions). Alternatively, if the
lower mass regions have less shielding from cosmic ray
 ionization than do high mass regions, the fractional
 ionization rate would be higher in the low mass regions, leading to an
 enhanced \HCO~abundance relative to \HCN~\citep{Marr1993}.

One significant region of~\HCN~emission without
significant~\HCO~emission is the linear filament, the swath of
\HCN~emission approximately 20\arcsec~west and 35\arcsec~north of
\sgra~(see location in Fig.~\ref{fig:h2hcn}).  We discuss this
feature in more detail below (\S\ref{sec:h2emission}) and argue
that the \HCN~abundance may be enhanced relative to \HCO~because of shocks
in the linear filament.  In
contrast, there is a single region where \HCO~emission dominates that
of \HCN. Between 17 and 30~\kms~in a region centered at 45\arcsec~east and
5\arcsec~south of \sgra, the median \HCO/\HCN~ratio is 1.22.  The
molecular gas cores found within this region tend to be some of the
lowest mass and density cores in the CND, supporting the argument of
higher \HCO~abundance relative to \HCN~in lower mass and density regions
due to enhanced cosmic ray ionization.  

\subsection{CND and H$_2$~Emission}\label{sec:h2emission}

In Figure~\ref{fig:h2hcn} we overlay our integrated \HCN~emission map
in contours upon a gray-scale image of the H$_2$~(1-0) S(1)
emission at 2.12$\mu$m from~\citet{Yusef2001}. H$_2$ can be excited by shocks or  
UV fluorescence.
The H$_2$ emission is most prominent in the CND in
the northeastern and southwestern lobes (as a likely result of limb
brightening) where the integrated \HCN~emission
is also the greatest.  \citet{Yusef2001} argue that the
most likely excitation sources for the H$_2$ emission in the CND are UV
radiation or the dissipation of energy from collisions
of dense molecular gas cores within the CND.  This core collision mechanism requires core 
densities of
$>10^6$~\cmcube, confirmed by our density measurements below
(\S\ref{sec:mass}).   Core collisions and/or shocks could also
create regions with low \HCO/\HCN~ratios, as shocks can help fuel
\HCN~emission and additionally increase \HCN~abundance relative to \HCO~\citep{Mitchel1983}.

The linear filament is
strongly detected in H$_2$ emission with two primary emission peaks
coincident with two \HCN~cores (cores X and Y in
Fig.~\ref{fig:corelocator} below).  This feature is extremely narrow
in H$_2$ ($\sim$1\arcsec) and contains an OH maser near
the northern end.  The emission may arise from shocks generated
by the supernova shell of Sgr A East hitting the
50~\kms~molecular cloud~\citep{Yusef2001}.  Such shocks may explain the extremely low \HCO/\HCN~ratio in the linear filament.

A similar situation exists for the outer filament, which begins at
the northern end of the CND northern arm.  Weak \HCN~emission is seen
coincident with the H$_2$ emission of the outer filament, but there is
no accompanying \HCO~emission.  We note that this emission is weaker
and does not
appear in contours in the integrated emission \HCN~map in
Figure~\ref{fig:h2hcn} but peaks around 10-15\arcsec~W, 90\arcsec~N.
In addition, the outer filament
coincides spatially and kinematically (peaking around 50~\kms) with [SiII] emission at 34.8 $\mu$m~\citep{Stolovy1997}.
The [SiII] emission and the \HCN~emission without accompanying~\HCO~flux both
suggest a shock excitation mechanism for H$_2$.

Two additional regions of good spatial agreement between the H$_2$ and
\HCN~are the northeastern and southern extensions. Both of these regions
are unshielded from the radiation of the central stellar cavity,
suggesting that the H$_2$ emission is caused by UV excitation.  We discuss the northeastern
extension in greater detail in Section~\ref{sec:ionized}.

\subsection{CND and the Minispiral}\label{sec:ionized}
To understand the relationship of the CND to the ionized minispiral, 
we
utilized two maps of the ionized emission.  The first, a {\it Hubble Space
Telescope (HST)} NICMOS Pa$\alpha$ map~\citep{Scoville2003} has an exquisite spatial
resolution of 0.2\arcsec.  The second, a Very Large Array (VLA) H92$\alpha$ radio
recombination line map~\citep{Roberts1993} combines 1\arcsec~spatial
resolution with 14~\kms~spectral resolution to provide kinematic
detail of the inner parsecs.

\begin{figure}[htb]
\begin{center}
\epsscale{0.9}
\plotone{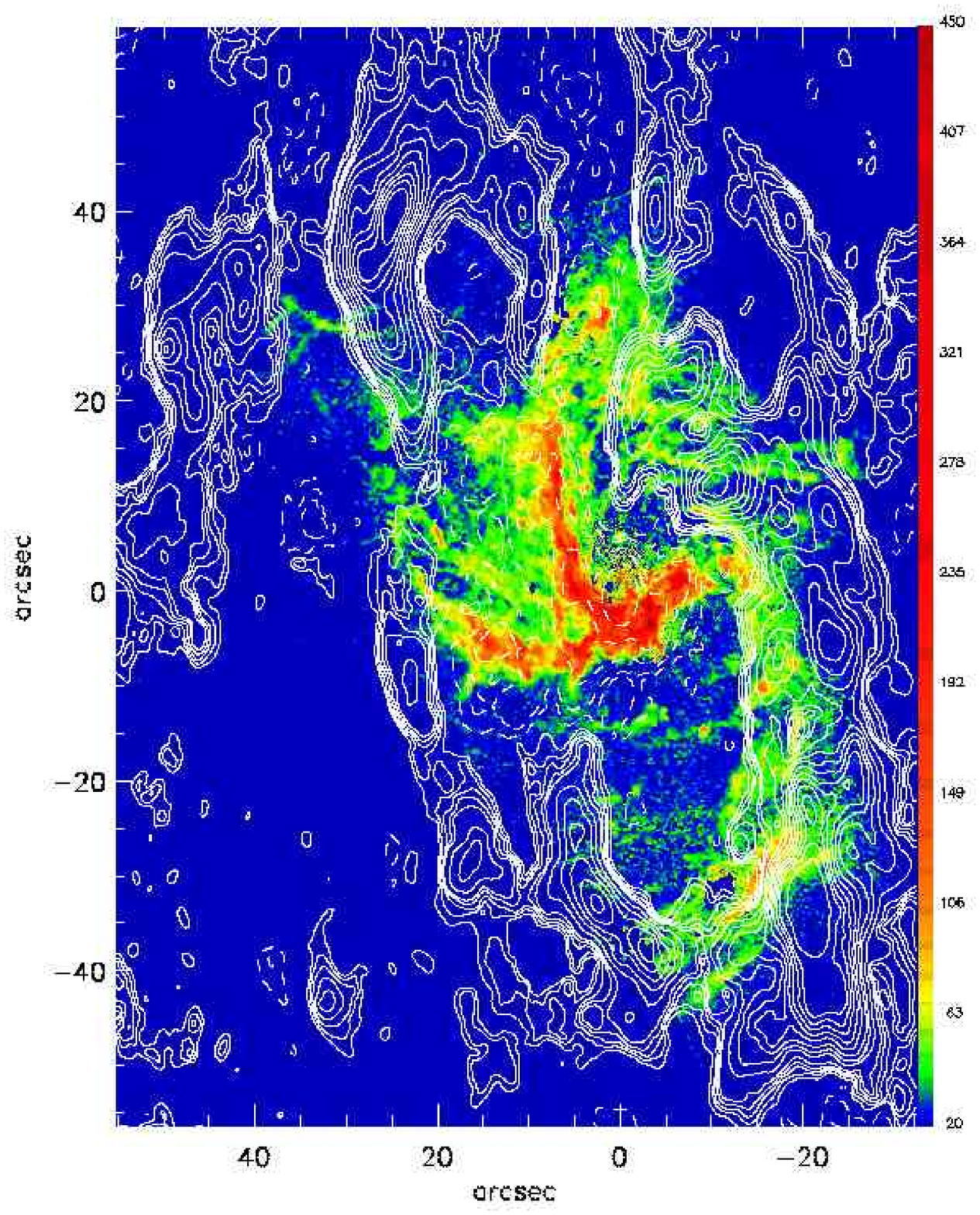}
\caption{Integrated \HCN~emission map in
  contours overlaid upon the color-scale {\it HST} NICMOS image of
  extinction corrected Pa$\alpha$ emission
  in the Galactic center region~\citep{Scoville2003}.
  Coordinates are offset from \sgra.  Emission contours (solid) are at
  [0.2, 0.4, 0.6, 0.8, 1.0, 1.5, 2.0, 2.5, 3.0, 3.5, 4.0, 4.5, 5.0, 6.0]*6.765~\jybeam~\kms~while the
  absorption contours (dashed) are at [-0.5, -1.0, -2.5, -5.0, -10.0]*6.765~\jybeam~\kms. The color-scale units, indicated on
  the right, are 10$^{-16}$~ergs cm$^{-2}$ s$^{-1}$ per pixel.}
\label{fig:hcnmoment0}
\end{center}
\end{figure}

\begin{figure}[htb]
\begin{center}
\epsscale{0.9}
\plotone{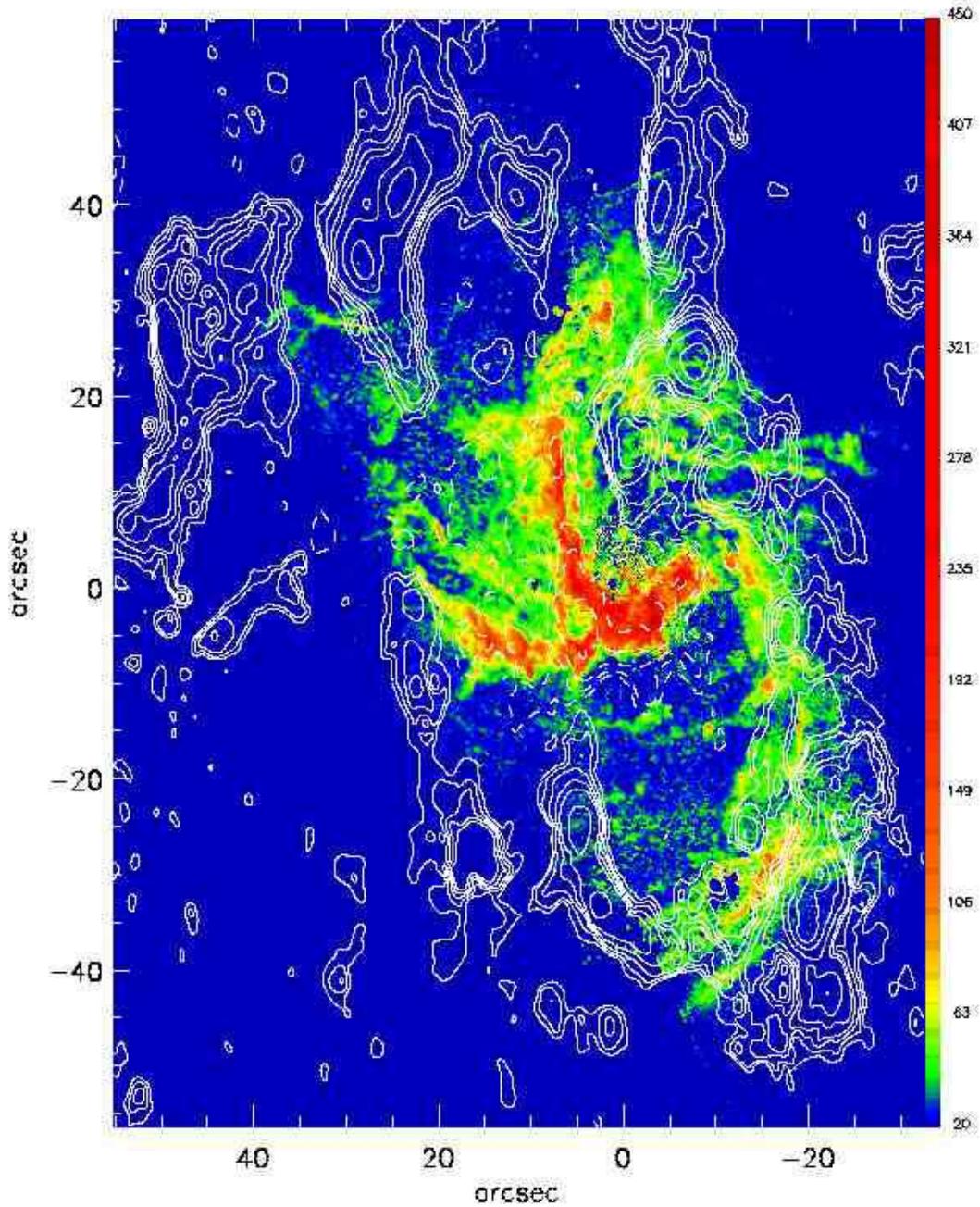}
\caption{Integrated \HCO~emission map in
  contours overlaid upon the color-scale {\it HST} NICMOS image of
  extinction-corrected Pa$\alpha$ emission
  in the Galactic Center region~\citep{Scoville2003}.
  Coordinates are offset from \sgra. Contour levels are the same as
  for the integrated \HCN~emission map (Fig.~\ref{fig:hcnmoment0}). The color-scale units, indicated on
  the right, are 10$^{-16}$~ergs cm$^{-2}$ s$^{-1}$ per pixel.}
\label{fig:hcomoment0}
\end{center}
\end{figure}

\begin{figure}[htb]
\begin{center}
\epsscale{0.75}
\plotone{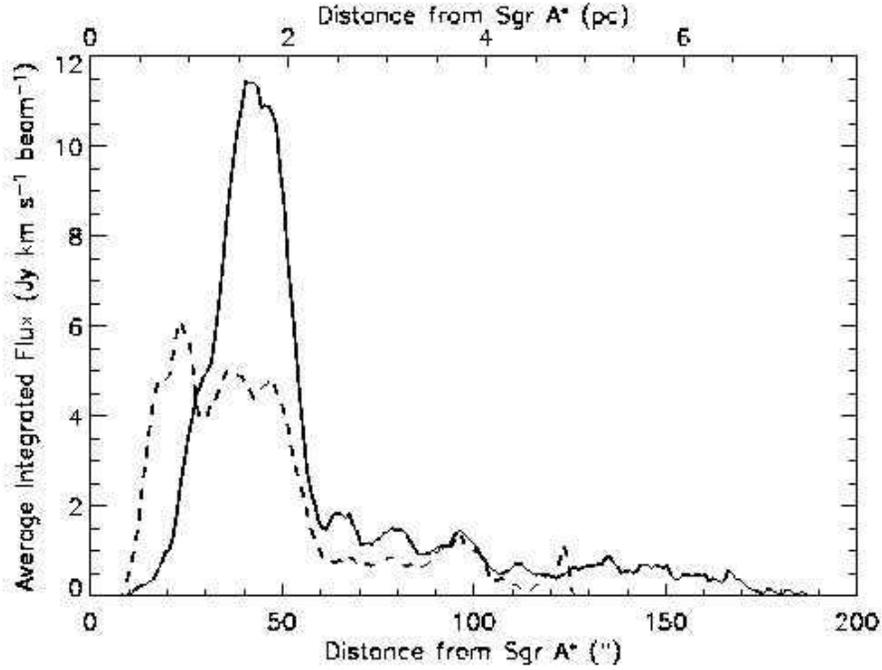}
\caption{Average integrated \HCN~flux per pixel as a function of distance from
  \sgra.  Distances are marked in both arcseconds (bottom) and parsecs
  (top).  The dashed line corresponds to the projected flux
  distribution, while the solid line is the deprojected flux
  distribution assuming an inclination angle of
  60$\arcdeg$~\citep{Jackson1993}. Note that the area under each of
  the graphs is not identical because different numbers of pixels fell
  in the various distance intervals for the projected and deprojected
  distributions.}
\label{fig:fluxversusdistance}
\end{center}
\end{figure}

\begin{figure}[htb]
\begin{center}
\epsscale{0.85}
\plotone{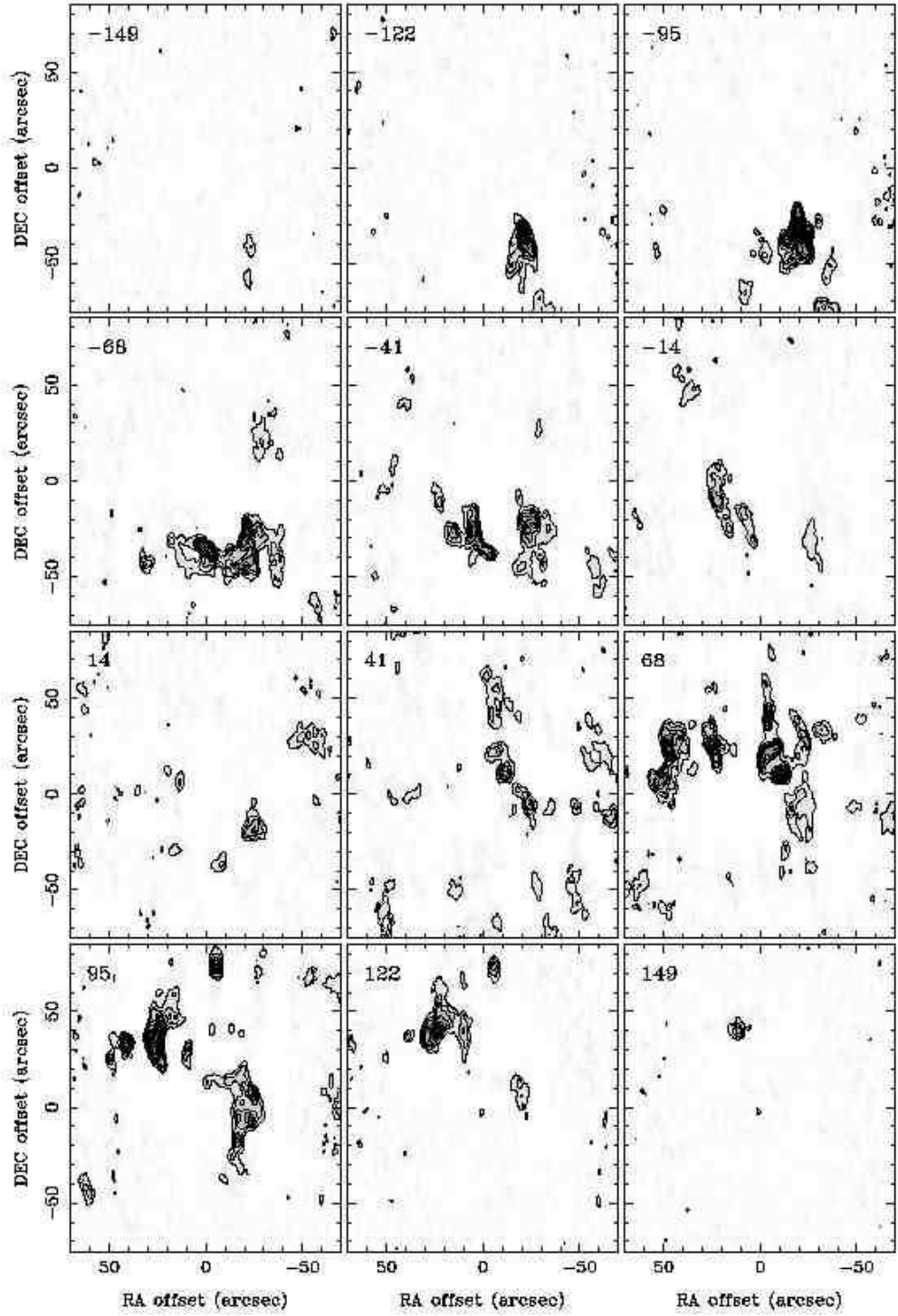}
\caption{\HCN~emission binned into 27~\kms~(four channel) intervals.  The lowest contour is at 5$\sigma$ (1$\sigma =
15$~mJy~beam$^{-1}$ for four channels) and each subsequent level
increases by 5$\sigma$. The central velocity in \kms~is listed in the
top left corner of each map.}
\label{fig:hcnchannels}
\end{center}
\end{figure}

\begin{figure}[htb]
\begin{center}
\epsscale{0.85}
\plotone{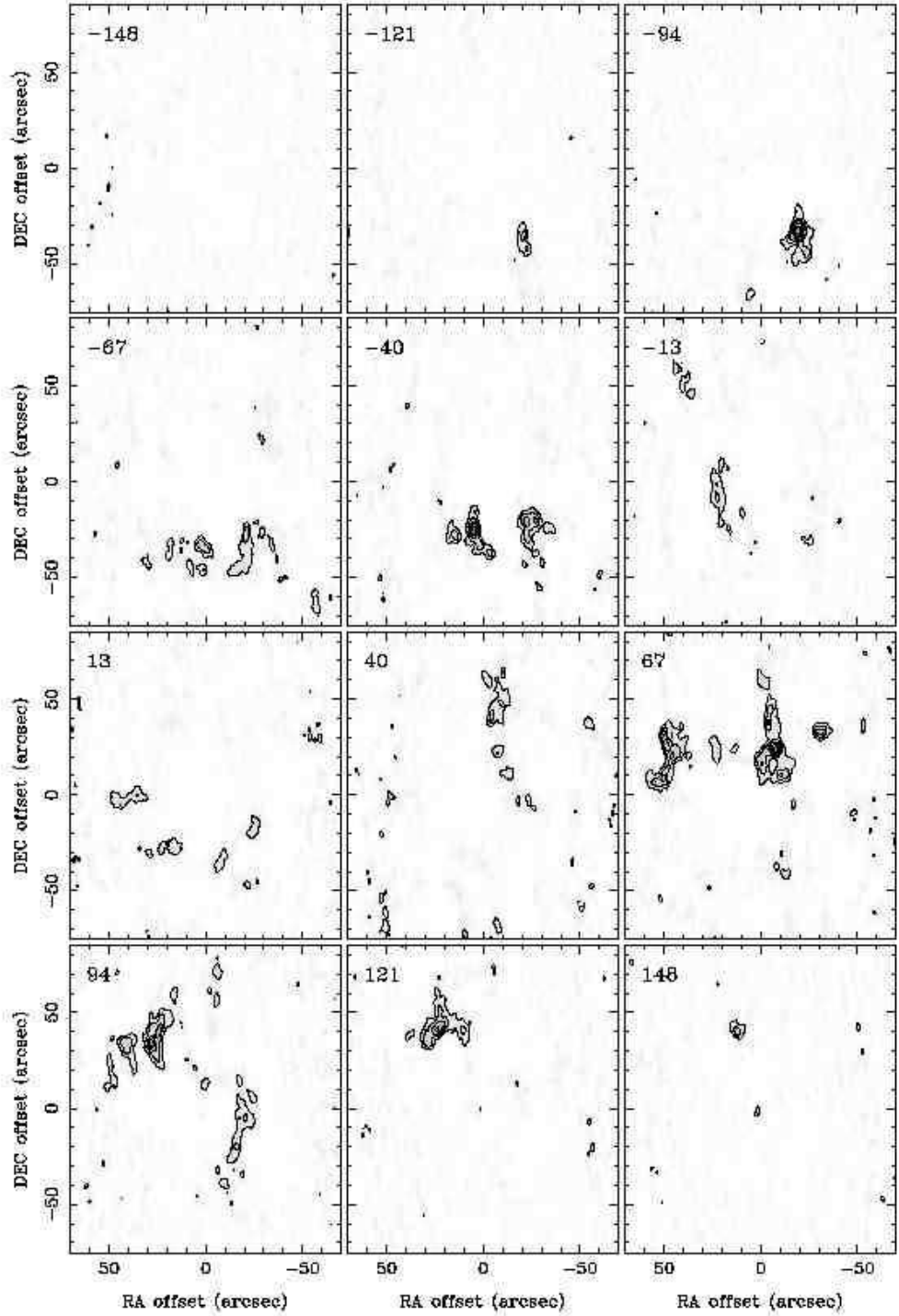}
\caption{\HCO~emission binned into 27~\kms~(four channel)
  intervals.  The contour levels are identical to the levels
in the \HCN~channel maps (Fig.~\ref{fig:hcnchannels}), so the
significantly reduced \HCO~emission relative to \HCN~is readily
  apparent.  The central velocity in \kms~is listed in the
top left corner of each map.}
\label{fig:hcochannels}
\end{center}
\end{figure}

\begin{figure}[htb]
\begin{center}
\epsscale{0.9}
\plotone{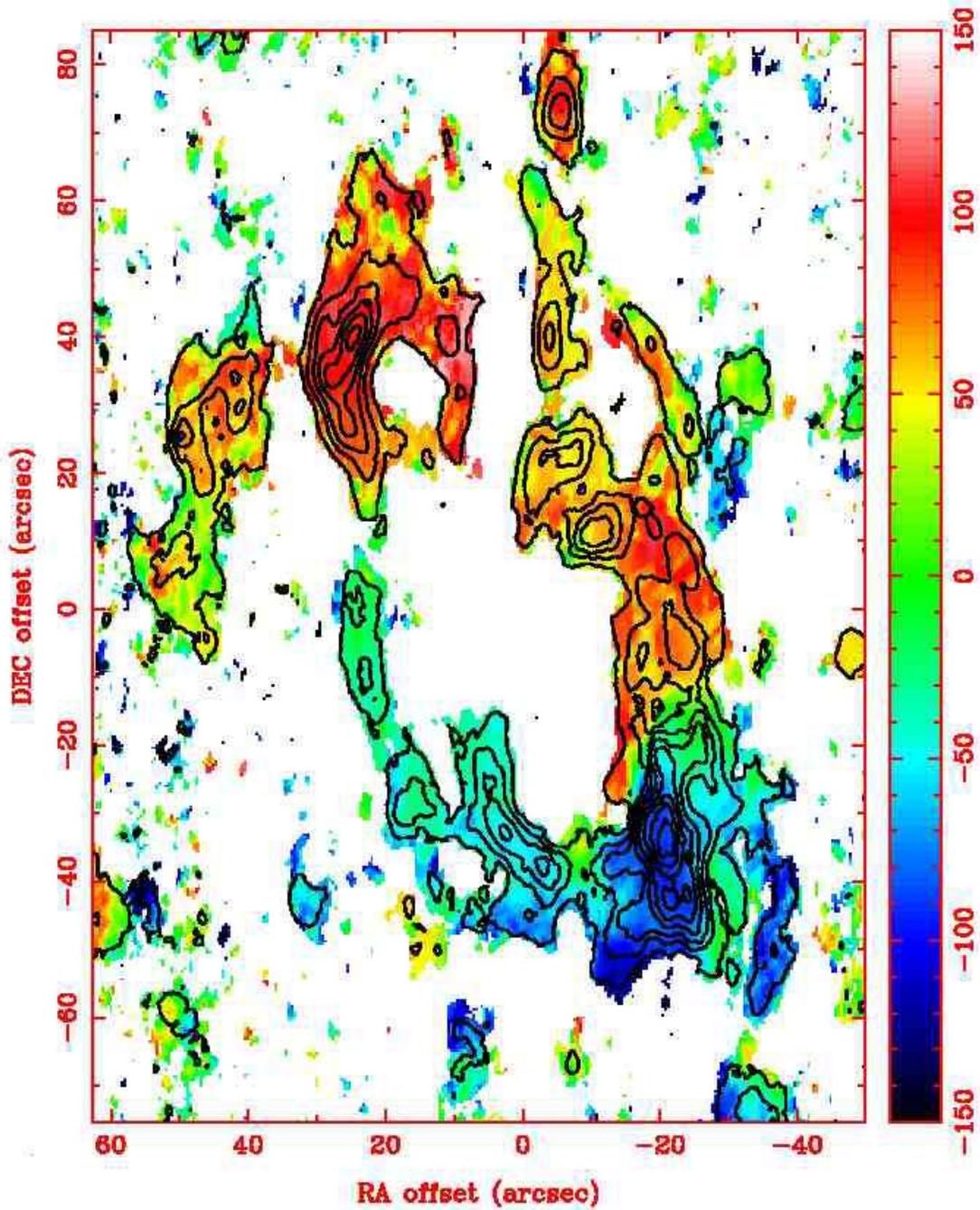}
\caption{
Flux-weighted average velocity (moment 1) map of \HCN~emission in color with
  integrated~\HCN~emission in contours.  Only individual channel
  emission at $>2\sigma$ is included.
  The velocities are with respect to V$_{\rm{LSR}} = 0$~\kms, and negative velocities (shown in blue) correspond to
  approaching emission. Emission contours begin at 3.4\jybeam~\kms~and
  are spaced by 6.8\jybeam~\kms.}
\label{fig:momentone}
\end{center}
\end{figure}

\begin{figure}[htb]
\begin{center}
\epsscale{0.9}
\plotone{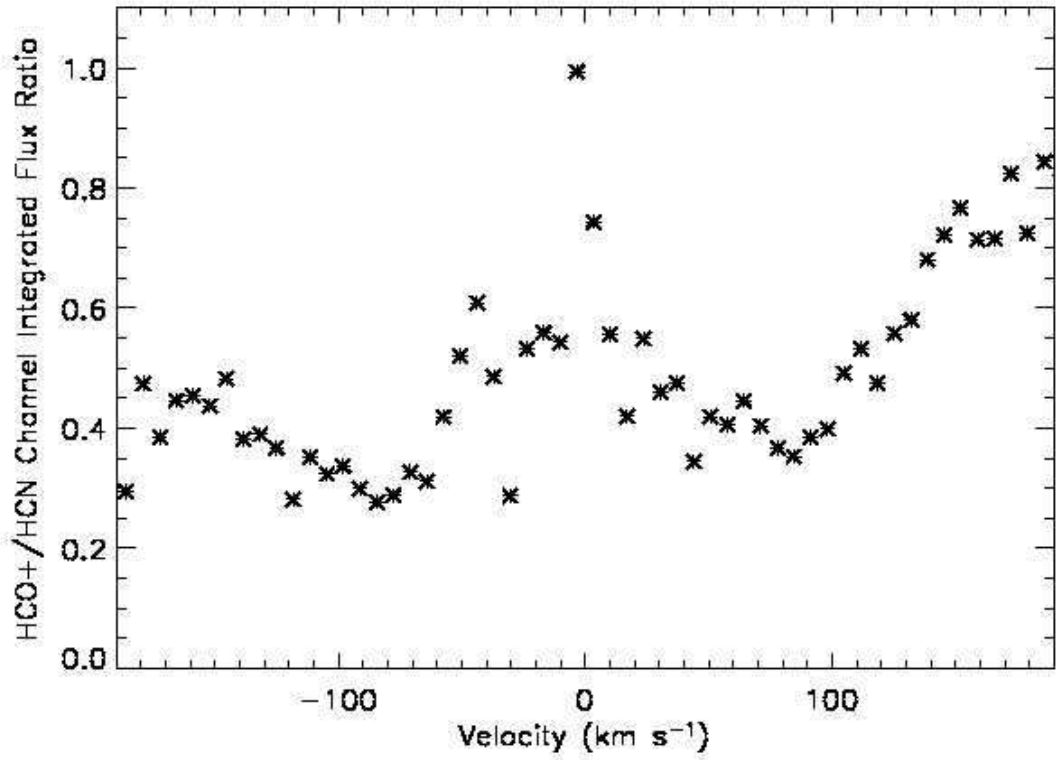}
\caption{Ratio of total \HCO~to total \HCN~emission in each channel as
  a function of velocity.  Only 3$\sigma$ positive
  emission was included in the flux ratio measurements.  The noise level is
  virtually identical for both the \HCN~and \HCO~maps.}
\label{fig:hcnhcoratio}
\end{center}
\end{figure}

\begin{figure}[htb]
\begin{center}
\epsscale{0.80}
\plotone{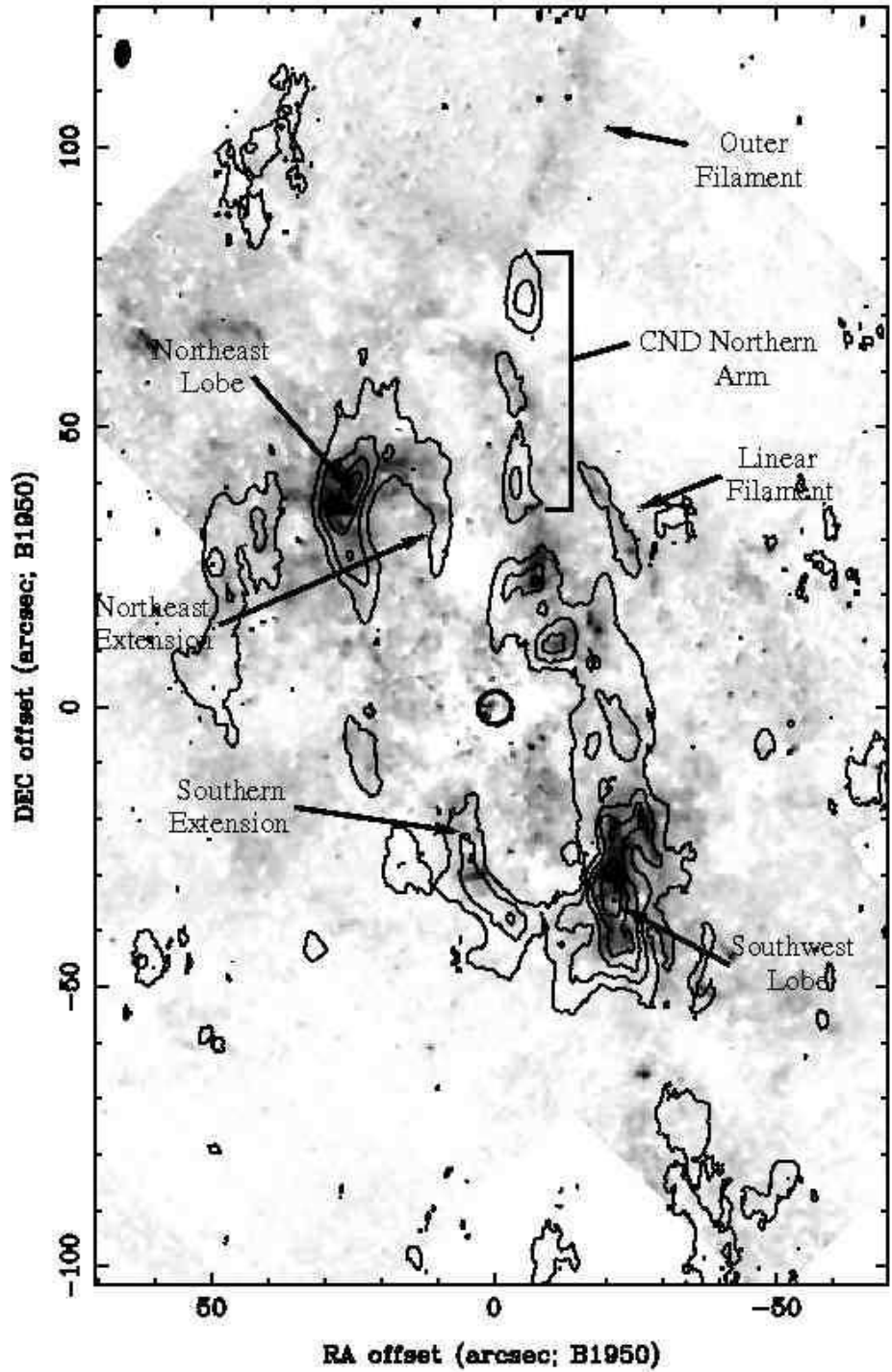}
\caption{\HCN~emission in contours overlaid upon a reverse gray-scale H$_2$
  (1-0) S(1) line emission map~\citep{Yusef2001}.  Significant features
  both in the H$_2$ emission and the \HCN~CND emission are labeled,
  and \sgra~is marked by a circle.  The contours begin at
  5.1 \jybeam~\kms~and are spaced by 10.1 \jybeam~\kms.}
\label{fig:h2hcn}
\end{center}
\end{figure}

\begin{figure}[htb]
\begin{center}
\epsscale{0.9}
\plotone{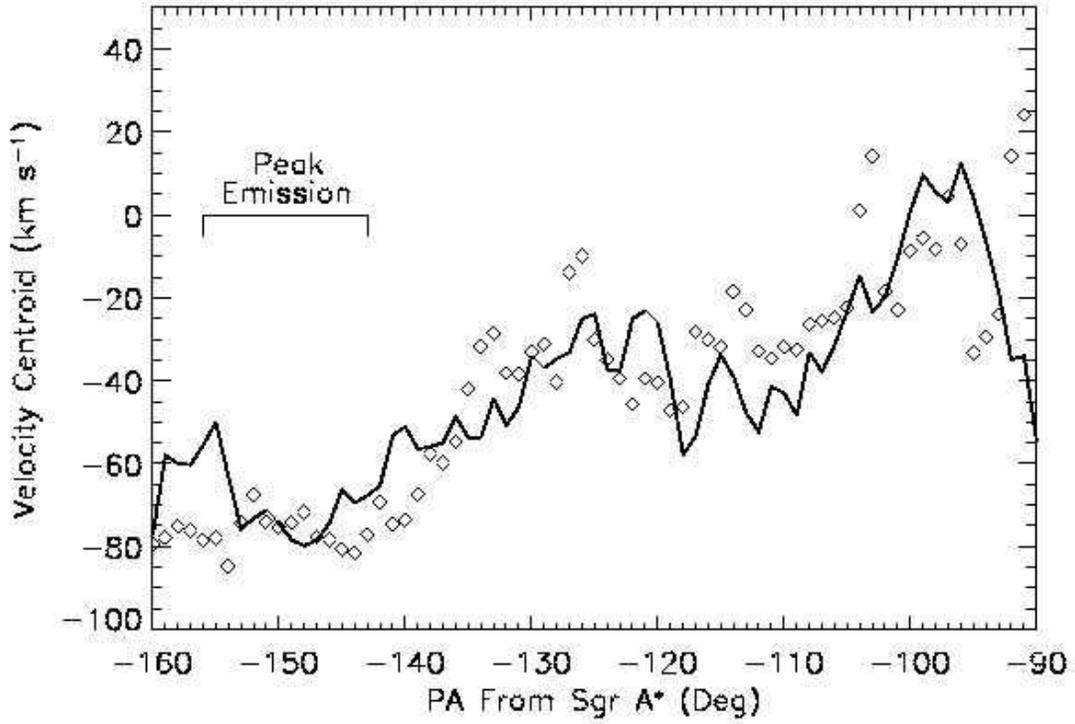}
\caption{Velocity centroids of \HCN~emission (line) and
  H92$\alpha$~emission (diamonds) along the western arc as a function of
  angle (degrees east of north) from \sgra.  The region of peak emission for both the
  ionized western arc and the western portion of CND, corresponding to
  Core O in Figure~\ref{fig:corelocator}, is indicated.}
\label{fig:h92hcnwesternarc}
\end{center}
\end{figure}

Figure~\ref{fig:hcnmoment0} shows our integrated \HCN~emission map in
contours on the color-scale Pa$\alpha$ image.  The most obvious
interaction is along the minispiral's western arc, where the ionized
gas is consistently immediately interior to the CND.  The western arc has been
suggested as the inner edge of the
CND, ionized by the strong UV radiation field of the stars in the
central cluster~\citep{Lo1983,Genzel1985,Serabyn1985,Gusten1987}.
We confirm this relationship kinematically through examining the velocity centroids of the ionized and
molecular emission along the western arc (Fig.~\ref{fig:h92hcnwesternarc}).  The velocities are in close
agreement, particularly at
a position angle of approximately~-145\arcdeg, at which both the \HCN~and H92$\alpha$~emission
peak.  While some overlap between
the western arc and the CND exists, the ionized emission always lies
interior to the peak \HCN~emission.  The high column densities that we
measure for the \HCN~emission along the western side of the CND
correspond to visual extinctions of up to 700
magnitudes~\citep{Scoville2003}, suggesting that in regions where
the ionized emission observed in Pa$\alpha$ overlaps the CND emission, the
ionized emission must lie near the foreground side of the molecular 
gas.

\begin{figure}[htb]
\begin{center}
\epsscale{0.9}
\plotone{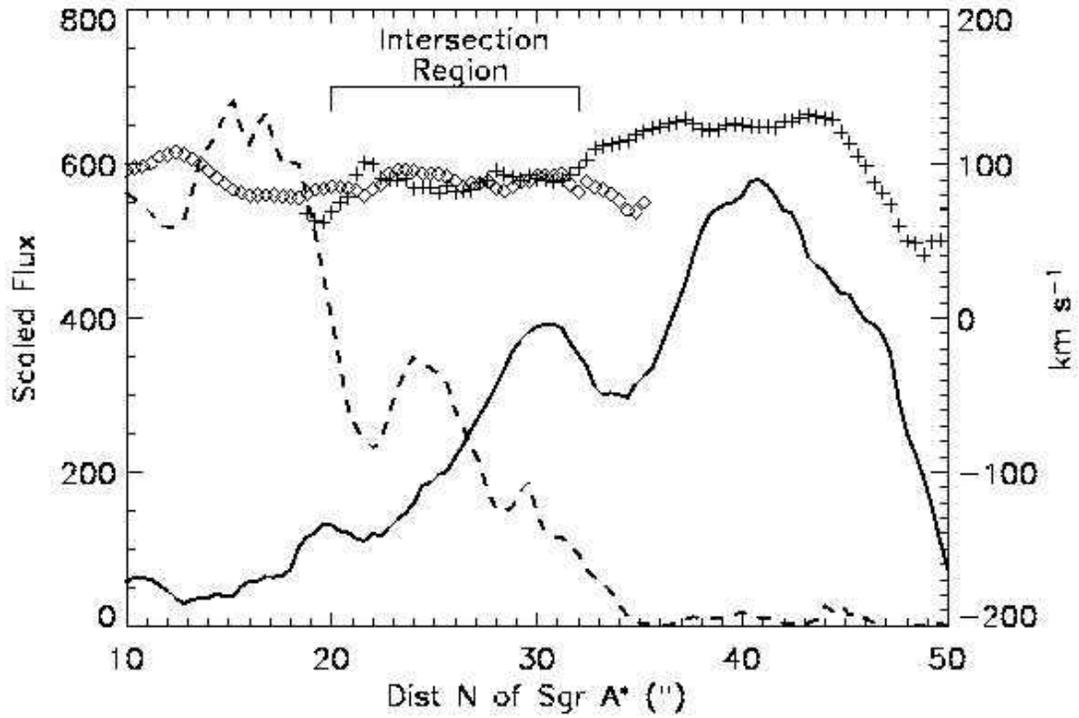}
\caption{\HCN~(solid line) and H92$\alpha$~(dashed line) scaled flux
  distributions along the minispiral northern arm and CND northeastern
  extension.  Velocity centroids are shown for the \HCN~emission (+'s) and
  H92$\alpha$~emission (diamonds).  The region of intersection between the
  northern arm and the northeastern extension is labeled.}
\label{fig:h92hcnnorthernarm}
\end{center}
\end{figure}

In addition, Figure~\ref{fig:hcnmoment0} suggests that the peak 
emission
from the northern arm of the minispiral may connect with the northeastern
extension from the CND at $\sim$~5\arcsec~E, 20\arcsec~N.  The
enhanced spatial resolution of our observations reveals the northeast
extension to be a longer and more sharply collimated feature than
previously observed.   The strong
collimation in the molecular and ionized gas can be explained by
pressure of mass-loss winds from emission-line stars in
the inner parsec~\citep{Scoville2003}.  At the 
intersection between the northern arm and
the northeastern extension the velocities are
in remarkable agreement (Fig.~\ref{fig:h92hcnnorthernarm}).  On the
other hand, the northernmost parts of the 
northern arm appear
to bend to the west slightly after meeting the northeastern 
extension and pass through the northern gap in the CND, and based on an 
increase in
CND linewidths on either side of the northern
gap and the presence of an OH maser within the
gap~\citep{YusefZadeh1999},~\citet{Wright2001} suggest that the
minispiral northern arm is infalling and creates the gap as it pushes
through the CND.  A definite conclusion on the relationship between
the northern arm and the CND is not possible with these observations,
but our improved spatial resolution allows us to suggest for the first
time that the northeastern extension and the northern arm may be part of
a single, highly collimated feature.

\section{Cores within the CND}\label{sec:coreproperties}
The enhanced spatial resolution of our \HCN~and \HCO~observations
resolves a great deal of internal gas structure within the CND, 
enabling us to analyze the molecular gas properties there using 
techniques similar to those used to derive the properties of Galactic 
giant molecular clouds (GMCs) and their cloud cores. 

\subsection{Core Identification and 
Measurements}\label{sec:coreidentify}
From the \HCN~maps we have identified 26 bright, isolated emission
features, each a dense molecular gas core within the CND.  These  features, labeled A-Z, are marked in
Figure~\ref{fig:corelocator}.  We required that each core included
in this analysis be isolated in both position and velocity space from
other emission peaks down to at least 50\% of the peak emission for
the core.  Included in this sample are the vast majority of the brightest
emission cores that met our requirements for isolation, as well as a
number of lower emission level cores; however, we
do not claim completeness to any particular flux level but rather
intend the sample to be representative of the range of bright cores found
within the CND.   Some peaks in the integrated \HCN~emission map
(Fig.~\ref{fig:hcnmoment0}), such as that 10\arcsec~west of core F, are too
close spatially and kinematically and thus are not included in our
core sample. There are additional emission cores at larger distances from \sgra~(e.g., 5\arcsec~W, 80\arcsec~N in
Fig.~\ref{fig:momentone}), but we restricted our sample to cores contained in the
main CND emission..  In Table~\ref{tab:clumpproperties} we detail the
properties of the 26 cores, including location, size, velocity center
and FWHM, and mass and density estimates.  Each of these core
properties is discussed in greater detail below.

For each of the cores, position-velocity strip maps were generated in
the north-south and east-west directions to use for measurements of core size
and velocity width.  In addition, the strip maps allowed us to reduce
line of sight blending and separate multiple kinematic components
(such as core pairs K/L and S/T).  Examples of these position-velocity maps
are given in Figure~\ref{fig:pvcuts} for cores B, I, J, O, and V.  In
each of these plots we indicate the adopted half-power size and
velocity FWHM estimates by the dashed lines. 
The measured apparent size FWHMs were deconvolved for the 
instrumental spatial resolution (beamsize) using the Gaussian approximation 
(i.e. $\rm{size}_{\rm{true}}^{2} = \rm{size}_{\rm{apparent}}^{2 } - 
\rm{size}_{\rm{instrumental}}^{2}$).  The mean core FWHM (corrected
for the instrumental resolution) is $8.8\times5.1\arcsec$ or $0.34\times0.20$~pc. 
Approximating the average diameter of each core by the geometric mean
of its north-south and east-west FWHMs, we find that the mean core diameter is
$6.7\arcsec$ (0.25 pc).

In Table~\ref{tab:clumpproperties}, we also include the projected and deprojected distance of each core 
from \sgra.  To deproject, we assumed an inclination angle of 60\arcdeg~for the
CND~\citep{Jackson1993}.  The median and mean deprojected core
distances are 1.67 pc and 1.80 pc respectively, and the distribution of
the 26 core distances follows very closely the deprojected flux
distribution found in Figure~\ref{fig:fluxversusdistance}.

\HCNF~has three hyperfine lines with relative intensities of 1:5:3 and
relative velocities of -7.07:0:4.8~\kms~\citep{Park1999}.  This
velocity separation is on
the order of the width of each channel (6.8~\kms).  We
modeled the increase in measured velocity FWHM due to the
hyperfine structure, as well as the instrumental velocity resolution
(taken to be one channel width) and found it to be a minor effect for velocity
FWHMs $>20$~\kms~but substantial for smaller
widths. The simulations were used to correct the measured velocity
widths of the CND cores.  We find median and mean velocity FWHMs for the 26
cores of 28.1~and 26.1~\kms~respectively, corresponding to
approximately 4 channels in our observations.   It should be noted
that the 1:5:3 intensity ratio of the three \HCNF~hyperfine lines was
derived for optically thin emission and so may not be the exact ratio
applicable in this situation, because the \HCN~emission is optically
thick.  However, comparison with linewidths measured from our
\HCO~observations for a smaller sample of cores suggests that the linewidths measured from the
\HCN~emission and contained in Table~\ref{tab:clumpproperties} are in
agreement with the \HCO~linewidths to within 15 - 20\%.

\subsection{Core Masses}\label{sec:masses}

We use three independent approaches to estimate or constrain the 
masses of the molecular gas cores in the CND: virial estimates
assuming the cores are gravitationally bound, an optically thick
estimate obtained by integrating the gas 
column density over the projected area of each core, and a lower limit
obtained under the assumptions that the \HCN~in the cores is optically
thin and the contribution from shocks to the \HCN~flux is minimal.
Each of these approaches has its own limitations (eg. the cores may 
not be in virial equilibrium and the HCN to 
H$_{2}$ abundance ratio is very poorly known for the CND).
In addition, the optically thin and optically thick estimates are
affected by the uncertainty in the role that
shock processing plays in producing the observed \HCN~emission.
Shocks, particularly from core-core collisions, could substantially enhance 
the HCN abundance and hence the observed HCN flux per unit mass of gas.  
If the true HCN abundance is in fact enhanced,  our estimates of the core masses and densities under both the
optically thin and optically thick assumptions would be
overestimated. The optically thin
calculations therefore can be considered a lower limit to the core mass and density
only under the assumption that the shock contribution to the
\HCN~emission is negligible.

Keeping in mind these assumptions and their possible effects, we find
that the virial and optically thick methods yield substantially higher
masses and densities for the CND than previously derived from lower
resolution \HCN~imaging.

\subsubsection{Virial Masses}\label{sec:vmass}

Virial mass estimates for the cores were computed from the measured 
sizes and velocity widths using 
\begin{equation}\label{eq:virial}
\sigma_{\rm{v}}^2 = \frac{3}{5}\frac{\rm{G M}}{\rm{R}}\Rightarrow M_{vir} = 200 \rm{R}_{\rm{pc}} \Delta \rm{V}_{\rm{km~s^{-1}}}^{2} ~~\rm{M}_{\odot}
\end{equation}
where R$_{\rm{pc}}$ is the radius (pc) of the core (taken as the geometric mean of
the deconvolved half widths in the north-south and east-west
directions) and $\Delta V$ is the 
velocity line-width (FWHM)
in \kms. We model the core as a uniform density sphere, introducing
the factor of~$\frac{3}{5}$ in eq.~\ref{eq:virial}.

\subsubsection{Molecular Column Density (\ta) Masses}\label{sec:cmass}

The core masses can also be estimated by integrating the 
molecular gas column densities over the surface area of each core.
 The optical depth of the observed HCN emission can be estimated from
the observed flux ratio of \HCN~and H$^{13}$CN emission. Using the
lowest resolution but highest signal-to-noise ratio data (L configuration tracks in
Table~\ref{tab:tracks}) of the \HCN~line and the additional L
configuration observation of the H$^{13}$CN line we obtain a mean \HCNF~optical depth ($\tau_{\rm
  {HCN}}$)~of 4 for the southwestern and northeastern CND emission lobes
(assuming a $C/^{13}C$ isotopic ratio of 30 for the Galactic 
center).  This measured optical depth is in excellent agreement 
  with previous estimates for the CND~\citep{Marr1993}. 

The total column density of H$_2$ can be derived from the 
\HCNF ~optical depth ($\tau_{\rm{HCN}}$) by: 

\begin{equation}\label{eq:thick}
\rm{N}_{\rm{H}_2} = \frac{\tau_{\rm{HCN}}}{\rm{A_{\rm{1-0}}}}\frac{8\pi\nu^3}{\rm{f}_{\rm{0}}~(\rm{g}_{\rm{1}}\rm{/g}_{\rm{0}})~\rm{c}^3}\frac{\Delta
  \rm{V}}{1 - \rm{e}^{\frac{\rm{h}\nu}{\rm{kT_{\rm{x}}}}}} = 1.00\times 10^{24} ~~{(\tau_{\rm{HCN}} / 4) ~\Delta \rm{V}_{\rm{km}~\rm{s}^{-1}} 
~[\rm{T}_{\rm{x}} / (50\rm{K})]^2 \over
\rm{X}_{\rm{HCN}} / 10^{-9} }  ~~\rm{cm}^{-2}
\end{equation}

\noindent where T$_{\rm{x}}$ is the excitation temperature of the HCN 
rotational levels, X$_{\rm{HCN}}$ is the abundance of HCN relative to 
H$_2$, A$_{\rm{1-0}} = 2.38\times 10^{-5}$ sec$^{-1}$~is the Einstein A coefficient for the HCN J=(1-0) transition
and f$_{0}=\frac{\rm{hB_0}}{\rm{kT_{\rm{x}}}}$, for which
B$_0$ = 44.316 GHz is the rotational constant for the \HCNF~transition.  The CND dust temperature is between 20 and 80
K~\citep{Becklin1982, Mezger1989}.  Since at the very high densities encountered in the CND, the dust and gas should be in 
thermal 
equilibrium, we have scaled the HCN excitation temperature to 50 K,
the midpoint of the dust temperature range.

 Integrating the $\rm{H}_2$ column density ($\simeq \rm{n}_{\rm{H}_2}$ 
2 R) over the surface area ($\pi \rm{R}^{2}$) of the core and multiplying 
by a factor of 1.36 to account for the expected He mass fraction within
the core, one obtains the 
total mass of gas  M$_{\rm{col}}$, 

\begin{equation}
\rm{M}_{\rm{col}} = 6.5\times 10^{2} ~(\rm{N}_{\rm{H}_2} / 10^{24}\rm{cm}^{-2}) ~\rm{R}^2_{0.1 \rm{pc}}  
~~\rm{M}_{\odot}
\end{equation}

The major uncertainties in the column density mass estimate lie in 
the adopted abundance of HCN relative to H$_2$ and in the excitation 
temperature. Here we have scaled the equations and used a value of 
$\rm{X}_{\rm{HCN}} = 10^{-9}$.  This value has been derived for a 
number of 
high density cores in Galactic GMCs \citep{Blake1987} and also for the 
CND by~\citet{Marshall1995} from analysis of multi-transition 
HCN observations. We should note that much higher values for
$\rm{X}_{\rm{HCN}}$ such as $8\times10^{-8}$ and $2\times10^{-8}$ have been
adopted by~\citet{Marr1993} and~\citet{Jackson1993},
respectively. However,~\citet{Marr1993} considered only T$_{\rm{x}} =
150 - 450~\rm{K}$, compared with 50 K adopted 
here based on the CND dust emission.  Previous works have taken a wide range of values for
T$_{\rm{x}}$ from 20 - 450 K (see Table 2 of~\citet{Jackson1993}),
although most atomic and molecular studies have taken a value of
T$_{\rm{x}}~>$~100 K. Since the derived column 
densities and masses scale as T$_{\rm{x}}^2$~X$_{\rm{HCN}}^{-1}$,
adoption of the higher HCN abundance and 
the higher T$_{\rm{x}}$ by~\citet{Marr1993} largely offset each other in their effect on the 
derived masses.  Shock processing of the molecular gas would also increase the
value of $\rm{X}_{\rm{HCN}}$, leading to an overestimate of the
core masses.

\subsubsection{Optically Thin 
Masses}\label{sec:thinmass}

Lastly, we obtain a lower limit to the core mass under the 
assumptions that shock-excitation is negligible and the HCN emission is optically thin and thus every spontaneous decay
from the J = 1 rotational level yields an 
escaping line photon. The observed line flux then translates directly 
into 
a total rate of \HCN~photons (p$_{\rm{photons}}$) by: 

\begin{equation}\label{eq:thin1}
\rm{p_{\rm{photons}}}=
\frac{\rm{L}_{\nu}\Delta\nu}{\rm{h}\nu} = \frac{\rm{S}\Delta\rm{V}(\frac{\nu}{\rm{c}}){4\pi\rm{d}^2}}{\rm{h}\nu}
\end{equation}

\noindent However assuming a rotational
temperature and HCN abundance, we can also express the total rate of
\HCN~photons by:

\begin{equation}\label{eq:thin2}
\rm{p_{\rm{photons}}} =
\frac{\rm{M}_{\rm{core}}\rm{X}_{\rm{HCN}}}{\rm{m}_{\rm{H}_2}} \rm{f}_1
\rm{A}_{\rm{1-0}}
\end{equation}

\noindent where m$_{\rm{H}_2}$ is the mass of a hydrogen molecule, and by equating eqs.~\ref{eq:thin1} and~\ref{eq:thin2}, 
calculate a minimum H$_2$ mass for each core.  The
optically thin assumption is clearly not true based on our
measurements of $\tau_{\rm{HCN}}~\simeq~4$~for parts of the CND, but
it does allow for an estimate of the lower limit to the mass (assuming
negligible shock contribution).  After correcting for 
the mass fraction of He, we obtain

\begin{equation}\label{eq:thin}
\rm{M}_{\rm{opt. thin}} = 28.6 ~~{\rm{S}_{\rm{Jy}} ~\Delta
  \rm{V}_{\rm{km}~\rm{s}^{-1}} ~(\rm{T}_{\rm{x}} / 50 \rm{K}) \over
(\rm{X}_{\rm{HCN}} / 10^{-9}) }  ~~\rm{M}_{\odot}
\end{equation}

\noindent where S$_{\rm{Jy}} ~\Delta \rm{V}_{\rm{km}~\rm{s}^{-1}}$ is the total emission 
line flux of the core. We have assumed a distance to the CND of 8
kpc. If there were significant shock processing, then the actual masses
and densities would be lower.

\subsection{Derived Masses and Densities}\label{sec:mass}

In Table~\ref{tab:clumpproperties} we list the estimated masses
(virial, \ta, and 
optically thin) for the 26 CND cores identified in
Figure~\ref{fig:corelocator}. In Figure~\ref{fig:coremasses} we plot
the distribution of these core masses.  In these calculations, we assume T$_{\rm{x}}=$ 50 K and X$_{\rm{HCN}} =
10^{-9}$.  The virial and \ta~mass estimates are in good agreement, ranging from 2000 to 1.36$\times10^5$ \mdot~for the selected cores with a median mass of 1.65$\times10^4$ and
2.43$\times10^4$ \mdot, respectively. The
lower limit, optically thin masses are typically a factor of 10 lower.
The total mass contained within the 26 cores is 5.5$\times10^5$, 7.90$\times10^5$, and 6.79$\times10^4$ \mdot~assuming
virial, \ta, and optically thin densities respectively.  The 26 cores
included in our sample cover a significant fraction of the dense
\HCN~emission from the CND; therefore we estimate a total mass within
the cores of the CND of 10$^6$~\mdot. We note
that~\citet{Shukla2004} estimated virial masses for a sample of nine cores in the
CND based on separate $7~\times~3$\arcsec~OVRO \HCO~imaging range from
$3.0\times10^3$~to $4.5\times10^4$~\mdot, similar to our virial mass estimates. \
The close
similarity between the virial and \ta~mass distributions
(Fig.~\ref{fig:coremasses}) supports the assumption that the
cores are in virial equilibrium.

The internal densities of the CND cores can be obtained from their 
mass estimates by:

\begin{equation}
\rm{n}_{\rm{H}_2} = 3.58\times10^6 ~~{ (\rm{M} / 1000~\rm{M}_{\odot}) \over
R_{0.1pc}^3 }  ~~\rm{cm}^{-3}
\end{equation}

\noindent where we again assume a uniform density sphere.   The
density estimates for each core under the three different mass
estimates are contained in Table~\ref{tab:clumpproperties}.  

We graphically show the derived core densities as a function of
deprojected distance from \sgra~in Figure~\ref{fig:coredensities}.   
The core densities are significantly larger
than the $10^5$~-~$10^6$~\cmcube~previously
estimated for the CND~\citep{Genzel1985,Marr1993}. Indeed our mean
virial and \ta~densities are 3.8$\times10^7$~and
5.0$\times10^7$~\cmcube, respectively.   
All of the optically
thin density estimates are~$>10^6$~\cmcube. Our density estimates are, in fact,
similar to those ($10^6 - 10^8$ cm$^{-3}$) derived
by~\citet{Jackson1993} from single dish, multi-transition HCN
measurements;  their 
estimates 
are obtained from analysis of molecular excitation and hence are 
entirely independent of the virial and column density (\ta) techniques used 
here. Similarly,~\citet{Shukla2004} find virial densities of
$4\times10^6$ to $3\times10^7$~\cmcube~for their sample of nine CND
cores.  As we will discuss later, the high densities of our large core
sample have
important implications
for core stability in the tidal field of \sgra~and for the possible role of
these cores in the formation of stars in the inner parsecs. 

A final independent check on our mass and density estimates of the CND is provided by millimeter and sub-millimeter
dust emission measurements from the
CND~\citep{Mezger1989,Dent1993}.~\citet{Mezger1989} detected the CND dust at $\lambda = 1.3$ mm. After subtracting 
substantial radio continuum from the ionized gas, they estimate a 
total mass 
of cold dust (at 20 K) and gas $\sim 2\times10^4$~\mdot~in the northeastern and 
southwestern CND lobes. In 450~$\mu$m James Clerk Maxwell Telescope (JCMT) observations at 7\arcsec~resolution, the flux of the CND is 
approximately 
100 Jy in a distribution that closely follows that of the
\HCN~observations of the CND (compare Fig. 3 in this work to Fig. 10a of~\citet{Dent1993}). For a mass opacity coefficient of 24 gr cm$^{-3}$ at
450~$\mu$m~\citep{Hildebrand1983} this implies a mass of $\sim 1\times10^4$ \mdot~for T$_d$ = 50 K. 
These estimates are more than a factor of 10 less than our
measurements of the molecular gas mass within the CND.  However, mass
estimates based on observed dust emission require assumptions for the dust opacity 
coefficient and dust-to-gas abundance ratio -- both of which could be 
quite different in the high density CND -- and the assumption that the
dust was optically thin at 450~$\mu$m.  We note that the typical
densities and sizes estimated in Table 2 for the cores yield typical
mass column densities of 50 to 100 gr cm$^{-2}$.  For standard dust
properties and abundances~\citep{Hildebrand1983}, these column
densities imply optical depths of $\sim 5$ at 450$\mu$m.  Thus the
observed 450$\mu$m fluxes may considerably underestimate the CND mass
because of opacity effects. Our 3 
mm continuum maps
show no evidence of dust emission associated with the CND molecular 
gas cores, but 
this is below our detection limit even for the very high 
column densities estimated above. 

One factor that could affect our measurements of the core masses and
densities is the resolving out of substantial extended \HCN~flux.
However, we compared our \HCN~maps to Nobeyama 45m \HCN~observations~\citep{Kaifu1987} and find that for the CND lobes we recover $\sim~75$\% of
the total \HCN~flux.

\subsection{Tidal Stability}\label{sec:tidal}

The tidal shear on CND cores from the gravitational
potential of \sgra~and the central stellar population is extremely
large.  The total mass enclosed as a function of Galactic radius (L)
can be modeled as:

\begin{equation}
\rm{M}_{\rm{G}} = 4.0\times10^6 +
1.6\times10^6 \rm{L}_{\rm{pc}}^{1.25}~\rm{M}_{\odot}
\end{equation}
This expression for total enclosed mass is taken
from~\citet{Vollmer2000} and modified to account for new measurements
of the mass of \sgra~\citep{Ghez2003a}.  

Therefore the mean internal density required for a core to be tidally 
stable is :

\begin{equation}
\rm{n}_{\rm{H}_{2}\rm{,tidal}} = 2.87\times10^7 ~~( \rm{L}_{\rm{pc}}^{-3} + 0.4\rm{L}_{\rm{pc}}^{-1.75} )  ~~\rm{cm}^{-3}
\end{equation}

\noindent We list the minimum densities for
tidal stability ($\rm{n}_{\rm{H}_{2}\rm{,tidal}}$) for each core in 
Table~\ref{tab:clumpproperties}. The mean value is $1.2\times10^{7}$ 
cm$^{-3}$. 

Comparing the calculated core densities with the minimum densities 
required
for tidal stability, we find that the vast majority of cores are dense 
enough to be tidally stable.  Indeed, all 26 cores have
\ta~densities sufficient for tidal stability, and 23 of the 26 cores
have sufficient virial densities.  We note that there is no correlation between measured core densities and deprojected
distance from \sgra~and that the virial and \ta~densities are often
3-5 times greater than the minimal tidal stability density
(Fig.~\ref{fig:coredensities}).  This is strong
observational support for widespread tidal stability and virial
equilibrium within the CND
cores because if the cores were 
being tidally disrupted, their {\it apparent} virial
densities would always approach the tidal stability limit as 
the cores gradually become tidally stripped.

\section{Discussion: Implications of CND Core Stability}\label{sec:implications}
\subsection{Lifetime of the CND}
The core densities that we measure for the CND are not only substantially
higher than previous estimates but also on average 3-5 times larger than 
necessary for tidal stability. Thus the lifetime of the CND cores may 
be considerably longer than their internal dynamical timescales of $\sim 10^5$ yrs. 

The ionized gas within the minispiral has been estimated at
100~\mdot~\citep{Sanders1998}.  We calculate a total molecular mass 
within
the CND cores of $10^6~\rm{M}_{\odot}$, giving an
$\frac{\rm{H}_{2}}{\rm{HII}}$ mass ratio of
~$\sim 10^{4}$.  Estimates for the
infall time of the ionized material 
are~$\sim10^4$~yr~\citep{Scoville2003,Vollmer2001b,Sanders1998}.  
If even
10\% of the molecular gas in the CND ultimately fell into the central
parsec and became the ionized material now visible in the minispiral, 
the mass ratio between the CND and the minispiral and the minispiral
infall time argue for a typical lifetime for the material in the CND of~$\sim10^7\rm{yrs}$.
 This time frame is a 
factor of
10-100 larger than previous estimates and is also
much longer than the dynamical/orbital time for the CND.  

The neutral, atomic gas contained in the inner parsec has been estimated at 300
M$_{\odot}$~\citep{Jackson1993}.  The bulk of this gas has been
modeled as an infalling atomic cloud with the northern and eastern
arms of the minispiral as the photoionized inner edges of this cloud.
If this atomic mass is included in the mass conservation arguments
above, this reduces the lifetime of the CND by a factor of 4, which
is still significantly longer than previous estimates.

The sharpness in the dropoff of the distribution of \HCN~flux and CND
cores as a
function of deprojected distance from
\sgra~(Fig.~\ref{fig:fluxversusdistance} and \S\ref{sec:coreidentify}) has implications for
the lifetime of the CND.  More than 70\% of the \HCN~emission from
our observations is contained within the inner 2 pc. If the CND has a 
$10^7$~yr
lifetime, as suggested from the mass conservation argument, then there
are sufficient molecular gas cores at radii beyond the CND that over $10^7$~yr could
migrate to the CND, providing a means of maintaining its mass.
~\citet{Coil1999,Coil2000}~find evidence for such feeding of the CND
from the 20~\kms~molecular cloud at a 10 pc projected distance from
\sgra and even detect cores within the streamer feeding the CND.  In contrast, a short lifetime for the cores of
the CND would suggest that the CND itself is a short-lived structure,
as there is insufficient additional material (as traced either by
the cores or by the raw \HCN~emission) outside the CND but within the
inner few parsecs for replenishment.  

\subsection{Formation of the Massive Young Stars Found in the Inner 
Parsec}\label{sec:formation}
Several stellar populations are observed within the central 
parsec~\citep{Genzel2001}.  These include a 1-10 Gyr stellar
population dominated by emission from red
giants~\citep{Genzel2003}, tens of blue supergiants tracing a very
recent (2 - 7 Myr) burst
of star formation~\citep{Forrest1987, Allen1990,
  Krabbe1991,
  Krabbe1995,Morris1996,Tamblyn1996,Blum1996,Paumard2001}, and a
population of presumably young, dust-enshrouded 
stars~\citep{Becklin1978, Krabbe1995, Genzel1996,
  Genzel2003}.~\citet{Genzel2003} describe the stellar population
of the inner parsec as a metal-rich older star
cluster, interspersed with a number
of young stars, which dominate the emission in the inner
10\arcsec. \citet{Scoville2003} detect a drop in the stellar surface 
brightness
within 0.8\arcsec~of \sgra, likely due to the interaction of the
black hole with the surrounding stellar environment.
   
Understanding the formation mechanism of the young stars within the 
inner
parsec has proven
extremely difficult.  The enormous gravitational potential of
\sgra~prevents in situ formation without the aid of shocks or
collisions unless molecular clouds have densities greater than
10$^9$~- 10$^{10}$~\cmcube.  While we measure high densities within the
cores of the CND, our measured densities are still 100-1000 times
below this in situ formation threshold.  On the other hand, star formation at
greater distances (ie. $>10$ pc) from \sgra~and subsequent inward
migration is also unfeasible because the time for two-body relaxation
of a star to the inner arcseconds is greater than the observed age of
the young stars.

Our discovery of the high densities and masses within the CND cores
provides a possible source for the young stars of the central parsec.
While the transport of individual stars to the central region is
impossible within the observed age of these young stars, star clusters can more quickly migrate to
the inner parsec via dynamical friction. \citet{Portegies2003} 
estimate that a $6\times10^4~\rm{M}_{\odot}$~or greater cluster with a diameter of 0.3
pc or less could reach the inner parsec from a radius of 5 pc or less
in the 2 - 7 Myr age of the young stars. This formation mechanism
had been previously dismissed in part because of the lack of massive
cloud cores that could produce such clusters in the inner 5 pc~\citep{Genzel2003}.  However, within our sample of 26
massive CND cores, almost all have diameters less than 0.3 pcs and
masses greater than 10$^4$~\mdot.  Indeed the most massive core has a
mass in excess of 10$^5$~\mdot.  These most massive cores could
potentially form a stellar cluster in the CND with high star formation
efficiency and migrate to the Galactic center within 2 - 7 Myr,
producing the observed young stellar population.  For example, evidence suggests that IRS13 may be a 
remnant of one such cluster~\citep{Maillard2004}.  In addition, simulations of the CND suggest that the CND
cores we observe are able to collapse and form stars before being shredded in 
the tidal field of the Galactic center (R. Coker 2004, private
communication).  Furthermore, numerical simulations have found that a
$10^5$~\mdot~cluster at the radius of the CND (or a less massive
cluster harboring an intermediate mass black hole) could spiral into the
inner parsec within the lifetime of the 2-7 Myr
stars~\citep{Kim2003,Kim2004,Hansen2003}.   Mid-infrared imaging of the CND~\citep{Telesco1996}
has revealed significant emission at 10, 20, and 30 $\mu$m,
especially along the western arc of the minispiral and more diffusely
over the entire western side of the CND.  There are no
significant peaks of mid-infrared emission coincident with the cores
visible in the CND.  Such a coincidence could have been an indication
of active star
formation.   Sensitive high spatial resolution observations in the mid
to far infrared (ie. 10-100$\mu$m) would allow for a more precise
search for ongoing star-formation within the CND cores.

Cores within the CND
reach densities of a few~$\times~10^7$~\cmcube, making them tidally
stable at distances less than 1 pc from the Galactic center (Fig.~\ref{fig:coredensities}).  An 
alternate explanation, therefore, for the young nuclear star
cluster is that a massive core remains
tidally stable for the majority of its infall from the CND onto 
\sgra~and forms stars
during this infall.  These new stars would then reach the inner parsec in less than 2 - 7 Myr.  Depending on the efficiency of the
star formation in the infalling core, multiple infall events might be
necessary.  These infall events could be spread over a period of a 
few Myr,
explaining the observed age spread within the nuclear stellar
cluster.

\section{Conclusions}\label{sec:conclusions}
We have presented OVRO millimeter observations of \HCN~and 
\HCO~emission from
the CND at a spatial resolution of~$5.1\arcsec\times2.7\arcsec$, a
significant improvement in spatial resolution over existing
observations.  With this enhanced resolution, we were able to study in
great detail the behavior of the CND in the gravitational potential of
the Galactic Center and the possible role of the CND in the
formation of the young stellar population of the inner parsec. In
particular we find:

1.  The CND, as traced by \HCN~and~\HCO~emission is a well-defined
    structure with an inner radius of $\sim40\arcsec$ (1.6 pc) from
    \sgra and a sharp drop past $\sim50\arcsec$~(2.0 pc),
    suggesting a ring-like morphology.  The main 
\HCN~and~\HCO~emission is consistent with rotation
    at $\sim110$~\kms, although emission at multiple velocities is
    found in some locations,
    particularly along the western edge of
    the CND. There is a lack of observed emission in the eastern part of the
    ring where radial velocities are near 0~\kms~because of strong 
line of
    sight absorption.

2.  The \HCN~flux is typically a factor of 2 - 3 greater than the
    \HCO~flux within the CND. However, there is substantial variation
    in this \HCO/\HCN~flux ratio with density, likely because of a higher fractional
    ionization from cosmic rays and thus greater abundance of \HCO~in
    lower density regions.  In addition, shocks may play a role in
    enhancing the \HCN~abundance and flux.

3.  There is good spatial agreement between the H$_2$ and \HCN~emission in the northeastern
    and southwestern CND lobes, as well as in the southern and northeastern extensions.  In these 
regions,
    the H$_2$ is likely UV excited or excited by collisions of the
    dense CND cores.  The linear and outer filaments are also seen significantly in 
H$_2$ and \HCN~(but not
    \HCO) and are likely shock-excited.

4.  There are a number of interactions between the ionized gas
    of the minispiral and the CND.  In particular, the western arc of
    the minispiral is both spatially and kinematically consistent with
    being the ionized inner edge of the CND.  In addition,, the
    minispiral northern arm may connect with the northeastern extension
    of the CND to form a single collimated structure.  

5.  We identify a sample of 26 resolved molecular gas cores within the
    CND.  These
    cores trace the overall \HCN~emission and have a
    characteristic diameter of~$\sim$~6.7\arcsec~(0.25 pc).  The cores 
have a typical density
    of (3-4)$\times10^7$~\cmcube, and a typical mass of (2-3)$\times10^4$~\mdot.  The total mass within
    the CND is estimated at $\sim 10^6~\rm{M}_{\odot}$.  Some of the
    HCN emission may be enhanced by shock chemistry, implying that our
    estimates of the optically thick and optically thin core masses
    and densities (assuming a standard 
\HCN abundance) are overestimates.

6.  The CND core densities are sufficient for tidal
    stability in almost all of the detected cores.  Based on mass
    conservation arguments, a 10$^7$~yr lifetime for the CND is predicted.  
 
7.  The core density estimates under the assumption of
    virialization and under the assumption of an optical depth \ta~are
    similar and often 3-5 times the minimum density for
    tidal stability, suggesting that many of the cores are indeed
    gravitationally bound.  

8.  The enhanced core densities and masses may explain the
    formation of the massive young (2-7 Myr) stars found in the inner
    parsec.  
The cores are tidally stable and can undergo
    star cluster formation while within the CND.  
    Alternatively, these cores are stable even at closer distances to
    \sgra~and could form stars during an infall process.
    In either scenario stars can be transported to the inner
    arcseconds within the observed age of the stars.
    
We thank Rob Coker, Sungsoo Kim, Milos Milosavljevic, and Farhad Yusef-Zadeh for
their discussions, comments, and insights on this project.  We thank
the anonymous referee for comments and suggestions that improved this work.
In addition, we thank Doug Roberts and Bill Dent for sharing their
data on H92$\alpha$~emission from the minispiral and 450 $\mu$m CND
emission respectively.  Research at the Owens Valley
Radio Observatory is supported by the National Science Foundation
through NSF grant AST99-81546.  M.H.C. is supported through NSF grant 
AST02-28955.


\begin{deluxetable}{c c l c c c c}
\tabletypesize{\small}
\tablecaption{Summary of Individual Observations\label{tab:tracks}}
\tablewidth{0pt}
\tablehead{
\colhead{Configuration} & \colhead{Largest} & \colhead{Date}
& \colhead{Usable} & \colhead{Source} &
\colhead{3c273} & \colhead{3c354.4}\\
 & Baseline & & Baselines & Integration Time & Flux & Flux \\
 & (m) & & & (minutes) & (Jy) & (Jy)\\
}
\startdata
L & 115.0 & 1999 Nov 13 & 15 & 114 & - & 7.0\\
  & & 2000 Apr 25 & 14 & 140 & 9.0 & 6.0\\
\hline
E & 119.3 & 2000 Mar 16 & 15 & 140 & 9.1 & 7.2\\
 & & 2000 Mar 17 & 15 & 136 & 9.1 & 7.2\\
\hline
H & 241.7 & 1999 Dec 17 & 15 & 140 & 9.8 & 8.5\\
 & & 2000 Jan 4 & 8 & 70 & 9.8 & 8.5\\
 & & 2000 Jan 5 & 15 & 139 & 9.8 & 8.5\\
 & & 2000 Jan 8 & 15 & 140 & 9.8 & 8.5\\

\hline
U & 483.3 & 2000 Feb 6 & 15 & 130 & 9.5 & 9.0\\
 & & 2000 Feb 8 & 15 & 120 & - & 9.0\\
\enddata
\end{deluxetable}

\clearpage

\begin{landscape}
\begin{deluxetable}{c c c c c c| c c | c c c| c c c c}
\setlength{\tabcolsep}{0.05in}
\tableheadfrac{0.01}
\tabletypesize{\tiny}
\tablecaption{Measured Core Properties\label{tab:clumpproperties}}
\tablewidth{0.0pt}
\tablehead{
\colhead{(1)} &  \colhead{(2)} & \colhead{(3)} &\colhead{(4)} &
\colhead{(5)} & \colhead{(6)} & \colhead{(7)} & \colhead{(8)} &
\colhead{(9)} & \colhead{(10)} & \colhead{(11)} & \colhead{(12)} &
\colhead{(13)} & \colhead{(14)} & \colhead{(15)}\\
\colhead{Core} & \colhead{RA} & \colhead{Dec} & \colhead{Proj.} & \colhead{Deproj.} &
\colhead{Size} &
\colhead{Cent.} & \colhead{Vel.} & \colhead{Virial} &
\colhead{$\tau_{\rm{HCN}}~=~4$} & \colhead{Opt. Thin} & \colhead{Virial} &
\colhead{$\tau_{\rm{HCN}}~=~4$} & \colhead{Opt. Thin} & \colhead{Min Stable}\\
\colhead{ID}& \colhead{Offset} & \colhead{Offset} & \colhead{Dist.} & \colhead{Dist.} &
\colhead{FWHM} & \colhead{Vel.} &
\colhead{FWHM} & \colhead{Mass} & \colhead{Mass} & \colhead{Mass} &
\colhead{Dens} & \colhead{Dens} & \colhead{Dens} & \colhead{Dens}\\
&\multicolumn{2}{c}{(\arcsec~from \sgra)} & \colhead{(pc)} &
\colhead{(pc)} &
\colhead{(pc)} & \multicolumn{2}{c}{(km/s)} & \multicolumn{3}{c}{($\times~10^3~\rm {M_\odot}$)} & \multicolumn{4}{c}{($\times~10^{6}~\rm {cm}^{-3}$)} \\
}
\startdata
A & 9.2 & 32.0 & 1.29 &1.34 & 0.17 & 107 & 30.9 & 16.55 & 15.04 & 0.90 & 91.60
& 83.23 & 4.98 & 18.83\\
B & 10.8 & 40.0 & 1.61 & 1.68 & 0.27 & 139 & 31.6 & 27.33 & 38.53 & 1.85 &
38.04 & 53.64 & 2.57 & 10.70\\
C & 24.0 & 40.8 & 1.84 & 1.86 & 0.34 & 105 & 43.7 & 65.63 & 83.82 & 8.24 &
46.37 & 59.22 & 5.82 & 8.30\\
D & 27.6 & 34.8 & 1.72 & 1.86 & 0.43 & 101 & 45.5 & 88.84 & 135.96 & 10.92 &
32.29 & 49.41 & 3.97 & 8.34\\
E & 25.2 & 26.8 & 1.43 & 1.62 & 0.22 & 78 & 30.8 & 20.92 & 24.32 & 2.85 &
55.97 & 65.06 & 7.62 & 11.59\\
F & 48.8 & 26.0 & 2.14 & 3.10 & 0.28 & 72 & 20.0 & 11.22 & 25.49 & 2.55 &
14.65 & 33.28 & 3.33 & 2.55\\
G & 50.0 & 8.8 & 1.97 & 3.42 & 0.29 & 65 & 13.1 & 4.98 & 18.05 & 1.80 & 5.76
& 20.87 & 2.08 & 2.05\\
H & 22.0 & -0.8 & 0.85 & 1.61 & 0.24 & -17 & 17.3 & 7.10 & 15.78 & 0.74 &
15.33 & 34.05 & 1.60 & 11.87\\
I & 22.0 & -10.0 & 0.94 & 1.87 & 0.26 & -18 & 15.0 & 5.77 & 16.12 & 1.23 &
9.68 & 27.06 & 2.06 & 8.15\\
J & 4.8 & -25.2 & 0.99 & 1.42 & 0.25 & -37 & 11.7 & 3.41 & 11.72 & 1.69 &
6.41 & 22.01 & 3.16 & 16.29\\
K & 2.4 & -32.8 & 1.28 & 1.67 & 0.25 & -71 & 24.5 & 15.27 & 25.67 & 2.06 &
26.76 & 44.99 & 3.60 & 10.81\\
L & 1.6 & -33.2 & 1.29 & 1.66 & 0.19 & -38 & 12.6 & 3.06 & 7.51 & 0.78 &
12.52 & 30.77 & 3.21 & 11.02\\
M & -3.2 & -37.6 & 1.46 & 1.70 & 0.26 & -64 & 19.0 & 9.49 & 21.47 & 2.29 &
14.78 & 33.43 & 3.56 & 10.28\\
N & -16.4 & -43.6 & 1.81 & 1.82 & 0.22 & -64.5 & 14.7 & 4.78 & 11.59 & 0.93 &
12.83 & 31.15 & 2.50 & 8.73\\
O & -21.2 & -34.4 & 1.57 & 1.60 & 0.33 & -108 & 36.5 & 43.96 & 64.58 & 9.23 &
35.04 & 51.48 & 7.36 & 12.04\\
P & -21.2 & -24.8 & 1.27 & 1.39 & 0.21 & -73 & 28.2 & 16.80 & 20.49 & 2.34 &
50.82 & 62.00 & 7.06 & 16.93\\
Q & -26.8 & -20.0 & 1.3 & 1.68 & 0.37 & -38 & 12.6 & 5.97 & 28.69 & 2.72 &
3.28 & 15.74 & 1.49 & 10.69\\
R & -18.0 & -7.6 & 0.76 & 1.16 & 0.14 & -37 & 12.2 & 2.02 & 3.66 & 0.24 &
22.91 & 41.63 & 2.77 & 27.06\\
S & -23.2 & -6.0 & 0.93 & 1.55 & 0.32 & 89 & 34.4 & 37.56 & 56.18 & 2.67 &
33.81 & 50.57 & 2.40 & 12.96\\
T & -24.0 & -5.6 & 0.96 & 1.62 & 0.24 & 44 & 33.5 & 26.39 & 30.08 & 1.45 &
58.23 & 66.36 & 3.19 & 11.73\\
U & -16.8 & -5.6 & 0.69 & 1.10 & 0.24 & 79 & 34.7 & 29.01 & 32.65 & 1.91 &
59.71 & 67.20 & 3.93 & 30.90\\
V & -10.8 & 10.8 & 0.59 & 1.13 & 0.19 & 58 & 24.2 & 11.27 & 14.59 & 2.42 &
45.11 & 58.41 & 9.70 & 28.95\\
W & -7.6 & 23.2 & 0.95 & 1.47 & 0.22 & 56 & 27.9 & 17.25 & 22.30 & 2.07 &
45.28 & 58.52 & 5.44 & 14.89\\
X & -24.4 & 27.2 & 1.42 & 2.67 & 0.16 & 64 & 28.4 & 13.04 & 12.05 & 0.75 &
88.59 & 81.85 & 5.12 & 3.57\\
Y & -18.8 & 39.2 & 1.69 & 2.82 & 0.21 & 78 & 35.3 & 25.90 & 24.88 & 0.94 &
91.92 & 78.71 & 2.97 & 3.15\\
Z & -3.6 & 40.4 & 1.57 & 2.08 & 0.24 & 58 & 39.6 & 37.63 & 37.00 & 2.37 &
78.22 & 76.91 & 4.93 & 6.33\\
\hline
\bf{Mean Core} & & & \bf{1.32} & \bf{1.80} & \bf{0.25} & & \bf{26.1} & \bf{21.20}
& \bf{30.70} & \bf{2.61} & \bf{37.92} &\bf{49.91} & \bf{4.09} & \bf{12.26}\\
\enddata
\end{deluxetable}
\end{landscape}
\clearpage

\clearpage

\clearpage

\clearpage

\clearpage

\clearpage

\clearpage

\clearpage

\clearpage

\clearpage

\clearpage

\clearpage

\clearpage

\clearpage

\begin{figure}[htb]
\begin{center}
\epsscale{0.9}
\plotone{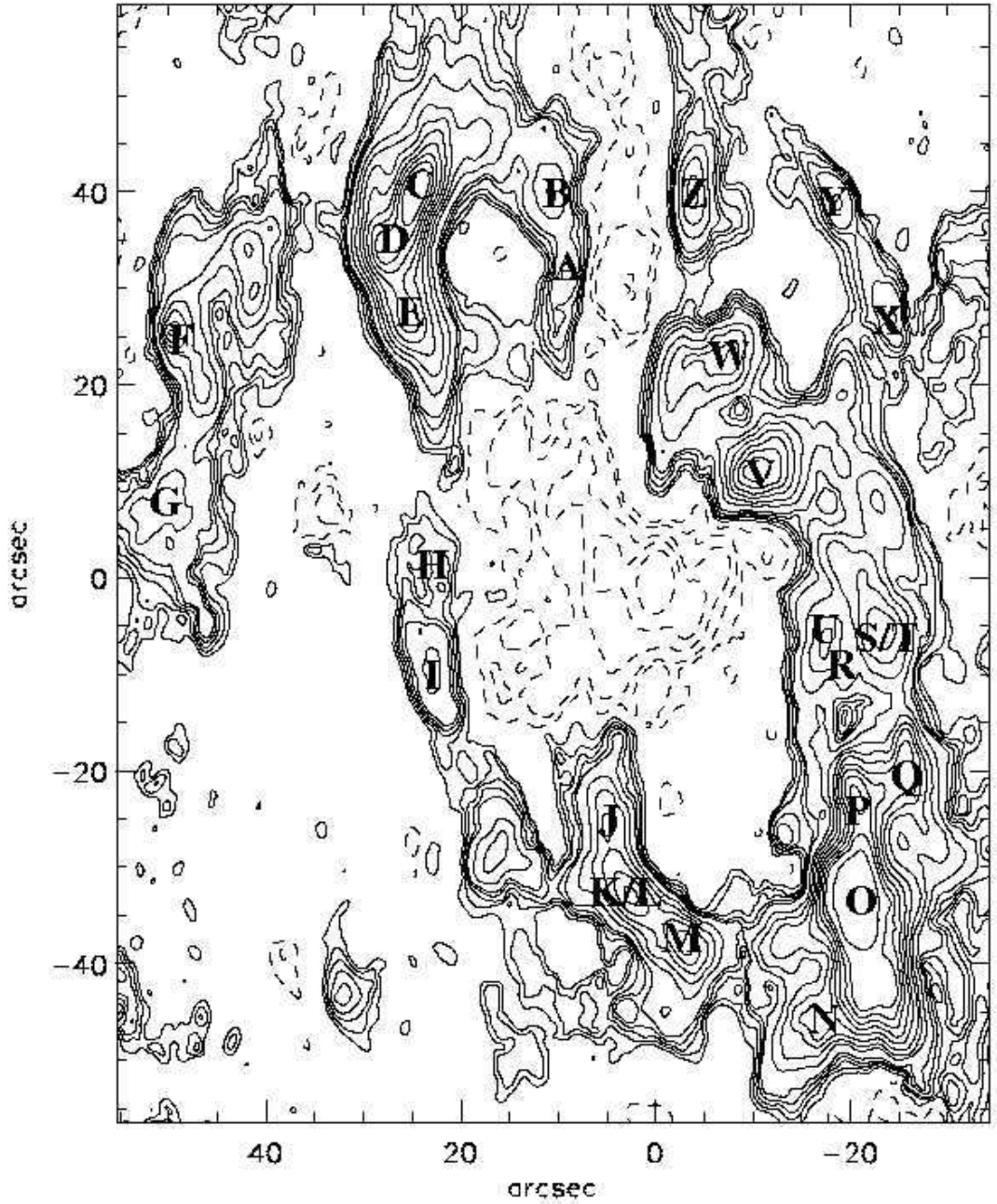}
\caption{\HCN~integrated emission contour map with locations of the
  isolated cores examined in Table~\ref{tab:clumpproperties}.  Cores
  K/L and S/T are pairs of cores at approximately the same spatial
  location but distinct in velocity space. The \HCN~contours are the
  same as in Figure~\ref{fig:hcnmoment0}.}
\label{fig:corelocator}
\end{center}
\end{figure}

\clearpage

\begin{figure}[htb]
\begin{center}
\epsscale{0.7}
\plotone{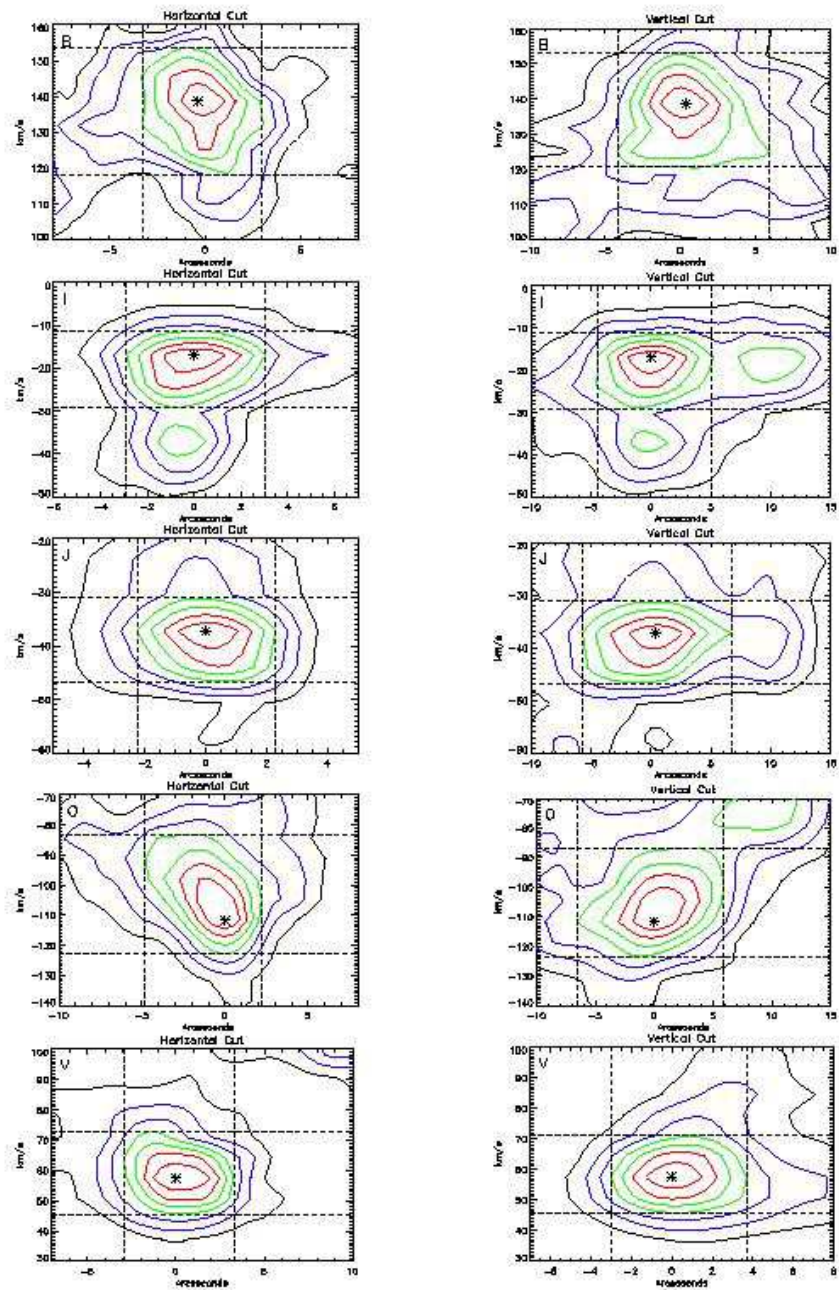}

\caption{Sample of position-velocity cuts in the horizontal (left) and vertical
  (right) directions centered on five cores identified in
  Figure~\ref{fig:corelocator}.  The horizontal and vertical cuts are summed over
  0.36\arcsec~and 0.2\arcsec~respectively.  The contours are
  set at 12.5\% increments of the peak intensity with the lowest green
  contour at 50\% and the lowest red
  contour at 75\%.  The asterisk
  indicates the location of peak intensity, and the dashed lines mark
  the region used for estimating the velocity and size FWHMs for each core.}
\label{fig:pvcuts}
\end{center}
\end{figure}

\clearpage

\begin{figure}[htb]
\begin{center}
\epsscale{1.0}
\plotone{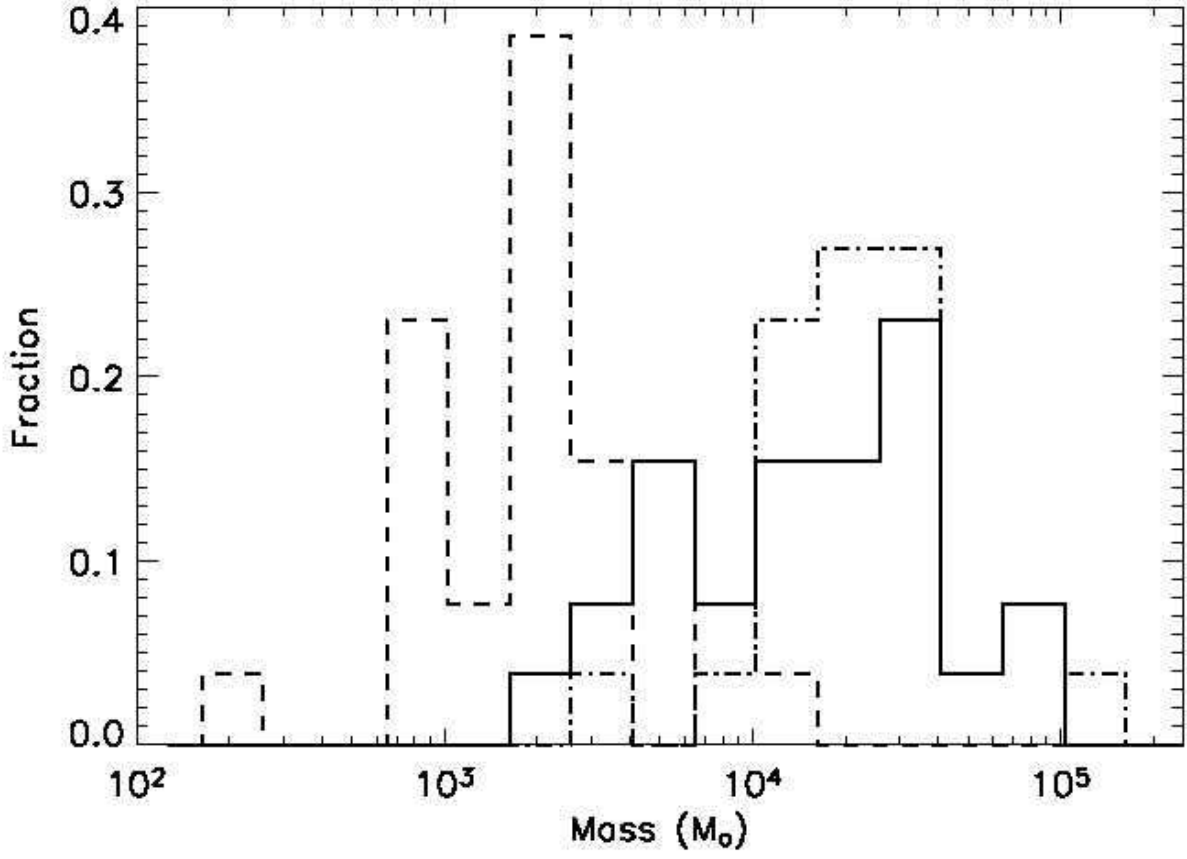}
\caption{Distribution of virial (solid), \ta~(dot-dashed),
  and optically thin (dashed) masses in logarithmic mass bins.}
\label{fig:coremasses}
\end{center}
\end{figure}

\clearpage

\begin{figure}[htb]
\begin{center}
\epsscale{1.0}
\plotone{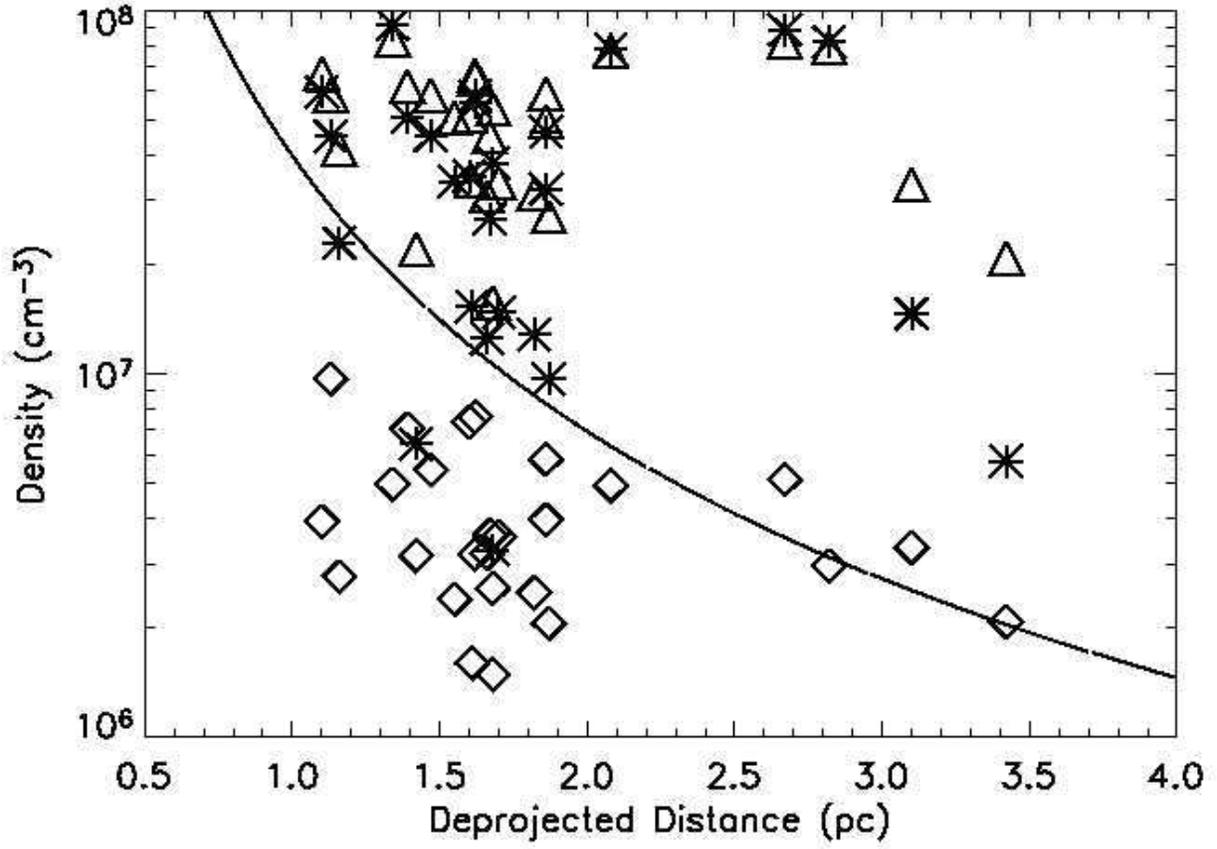}
\caption{Distribution of virial (asterisks), \ta~(triangles),
  and optically thin (diamonds) densities as a function of deprojected
  distance from \sgra.  The minimum density for tidal stability is
  shown by the solid curve.}
\label{fig:coredensities}
\end{center}
\end{figure}

\end{document}